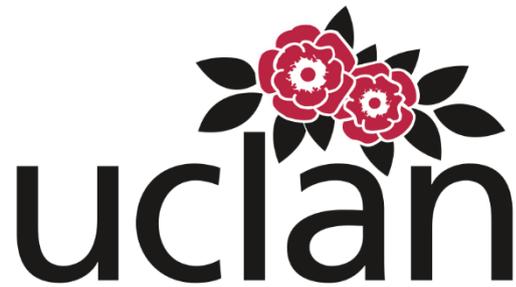

University of Central Lancashire

School of Computing, Engineering & Physical Science

Preston, United Kingdom

---

# Astrophysical Nuclear Reactions: from Hydrogen Burning to Supernovae Explosions

---

*Author*: Francesco Cuda                    *Supervisor*: Dr. Alex Dunhill

*A dissertation submitted in partial fulfilment*
*of the requirements for the degree of*
*Bachelor of Science (Hons) in Astronomy*

Submitted: 25 April 2018

Student ID: G20508157

Word count: 7941

## Assessment committee:

*Dr. Megan Argo*

*Dr. Alex Dunhill*

Jeremian Horrocks Institute for Mathematics, Physics and Astronomy

University of Central Lancashire, School of Computing, Engineering & Physical Sciences, Preston, United Kingdom

# Preface

This dissertation is the accomplishment of my Bachelor (Hons) degree in Astronomy at the School of Computing, Engineering and Physical Science, University of Central Lancashire, Preston, UK. The work leading to this dissertation was carried out between October 2017 and April 2018 under the supervision of Dr. Alex Dunhill. The achievement of this dissertation has been possible thanks to many people that have expired and helped me. First of all I would like to thank you my dissertation supervisor Dr. Alex Dunhill for the meticulous guidance during the preparation of this study and his constructive comments. I would like to thank the Course leader Dr. Barbara Jane Margaret Hassall for her support, professionalism and advice during this years-long journey at UCLan. I would like to thanks also my personal tutor Dr. Silvia Dalla that thrusted me since the beginning. Furthermore, I would also like to extend my gratitude to module tutors Dr. Roger Clowers, Dr. Anne Sansom, Dr. Jason Kirk and previous UCLan staff members Dr. Simon Murphy for their help and feedback.

.

# **Contents**

# Abstract

The work reported in this dissertation concerns the study of kinetics, cycles and quantum mechanic processes of nuclear reactions involved in low-mass stars such as our Sun and massive stars together with explosive nucleosynthesis phenomena having a key role in supernovae explosion. The first chapter provides an overview of the background of the nuclear physics and of the potential approaches to the field as quantum mechanic aspects, cross sections and the Maxwell-Boltzmann distribution. The next chapter describes the hydrogen burning processes for low mass stars (i.e. our sun), the CNO and hot CNO cycles that dominate in higher mass stars, and other cycles such as sodium and magnesium. The third chapter reviews helium burning process active in Red Giants as a result of the exhausted hydrogen fuel. A short overview of other helium processes is briefly discussed. The fourth chapter provide a summary of the accepted model of explosive burning process involved in supernovae explosion. Explosive nucleosynthesis together with heavy elements as silicon burning processes are also described. Appendices are also included containing mathematical treatments and details of the Laboratory for Underground Nuclear Astrophysics (LUNA) result measurements for different nuclear reactions.

*"Whereas all humans have approximately the same life expectancy the life expectancy of stars varies as much as from that of a butterfly to that of an elephant."* (George Gamow 1938)

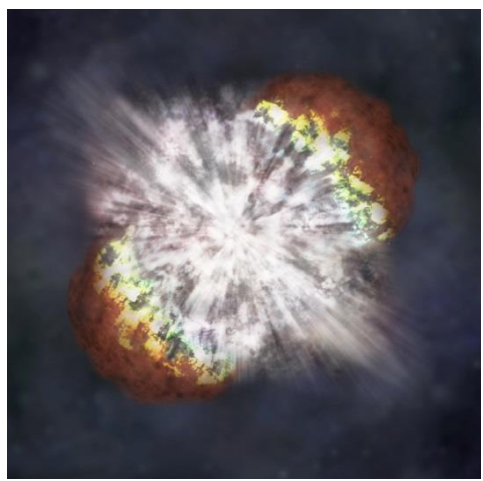

Figure A Supernova explosion (image credit NASA 2013)

# Introduction

## Background of dissertation

Nuclear astrophysics is one of the most active and intriguing interdisciplinary field in modern science that combines stellar evolution and nucleosynthesis studies. For years the exact mechanism by which stars generate energy to shine was not understood until 1920 when Aston determined the helium mass and successively Eddington suggested that the sun generate energy converting hydrogen into helium. Despite that Eddington could not justify the nuclear fusion from the observed stellar temperature. In 1928 Gamow gave the nuclear physics bases introducing the quantum mechanical probability using the tunnel effect through potential barrier explaining also the alpha-particle decay phenomenon. The first experimental nuclear reactions using an artificial collider disintegrating lithium nuclei with protons accelerated was used by Cockcroft and Walton in 1932. This experiment involving lithium disintegration in two alpha particles gave the physical bases for *pp-chains*. In 1934 Lauritsen and Crane were able to generate a 10min radioactivity bombarding carbon nuclei with protons delivering the first measurement later called *CNO cycle*. Hoyle in 1946 published the first stellar evolution framework using nuclear data available, but at the time experiments suggested that no stable nucleus of mass number 5 or 8 could exist in nature. This created a massive dilemma in the scientific community solved by Salpeter in 1951 that suggested the conversion of the unstable $^8$Be nucleus into the stable $^{12}$C by capturing another alpha particle and generate the necessary energy source in red giant stars (*triple-α reaction*). Since then astrophysical nuclear field has been evolved massively linking several different topics from astronomical observation, experiments, nuclear and stellar evolution theories. In addition, the advent of space era such as the Hubble Space Telescope (HST) has increased the scientific interest about star formation and the nuclear physics involved. Interstellar gas clouds are the nest of star formation from which they emerge following several stages as shown in Figure B:

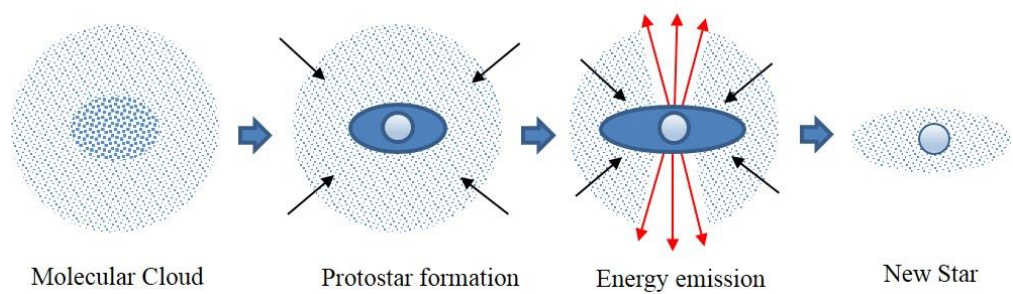

Molecular Cloud        Protostar formation        Energy emission        New Star

Figure B Star Formation stages

A giant molecular gas cloud having thousand times the solar mass contracts. When the cloud reaches a density value corresponded to $10^{-16}\,\mathrm{kg\,m^{-3}}$ a fragment is able to contract independently forming clusters of primitive stars called protostars. A protostar dominated by the opacity and ionised interior slowly contract controlling the rate of the releasing energy lost from the surface and gravitational energy. As consequence of that the temperature and pressure inside the protostar increase reaching internal temperature of 10 million K or higher enough to trigger thermonuclear reactions and the fusion of hydrogen to helium.

Thus, aim of this dissertation is to discuss and analyse astrophysical nuclear reactions involved in low-mass stars (i.e. our Sun) and massive stars from hydrogen burning to supernovae explosion neglecting other nuclear astrophysics features as Type Ia Supernovae, big-bang nucleosynthesis and burning processes beyond iron. Recent nuclear physics experiments and new simulation techniques allow analysis of the fundamental processes involved in nucleosynthesis. In the last decade the reaction sequence of the cold and hot CNO cycles have been re-analysed providing additional data to further understand the stellar hydrogen burning mechanism. Additionally, other advanced nuclear physics experiments in similar terrestrial accelerator and collider centres have modified current astrophysical theories, refining existing error percentage data. Thus, key questions to address are: how do stars produce energy? How do massive stars evolve and why do they explode as supernovae? How do these new findings give rise to existing astrophysical nuclear theories?

# CHAPTER 1: Thermonuclear Reactions

## 1.1 Nuclear Energy Source

Thermonuclear reactions are the engines that allow stars to shine and the trigger of stellar nucleosynthesis. These processes provide the necessary energy to compensate for the luminosity lost from the star's surfaces and they recycle chemical elements in the interstellar medium to create new generation of stars and planets (Weiss 2008) Figure 1:

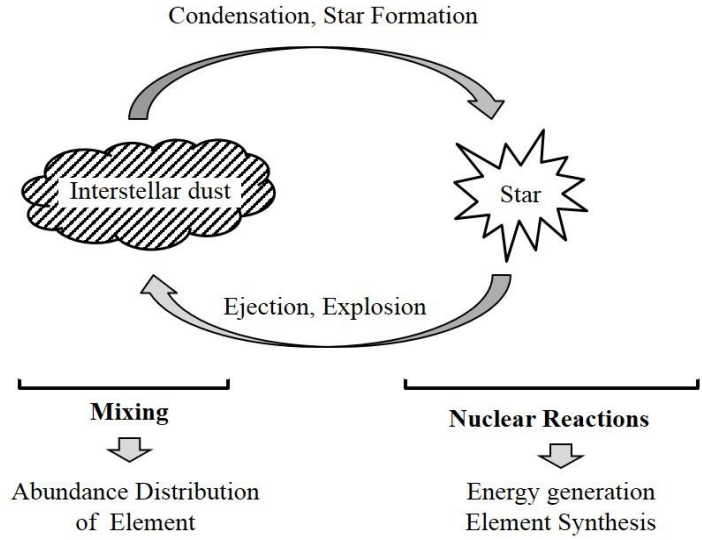

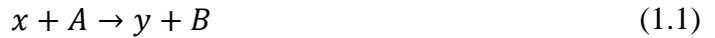

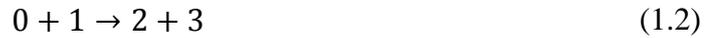

Figure 1 Cycling process between interstellar medium and stars

The main elements that fuel stars are hydrogen and helium. The production of nuclear energy proceeds through a sequence of reactions until complete conversion in iron occur. Nuclear reactions can be represented symbolically as:

$$x + A \rightarrow y + B \qquad (1.1)$$

or alternatively by numbers:

$$0 + 1 \rightarrow 2 + 3 \qquad (1.2)$$

where the (1.1) can be also indicated as $A$(x, y)$B$. Each possible combination of these numbers is called partition and can have different state of excitation named "The Reaction Channel". On the left side, equation (1.1) and (1.2), $x$(0) is the projectile and $A$(1) the target nucleus defined as entrance channel while $y$(2) and $B$(3) are the emerging nuclei or exit channel. Applying the mass-energy conservation law, the nuclear reaction $Q$-value can be written:

$$Q = (m_0 + m_1 - m_2 - m_3)c^2 \qquad (1.3)$$

where $m_0$ and $m_1$ are the nuclei masses in the entrance channel; $m_2$ and $m_3$ in the exit channel. For positive value of Q energy is produced so the reaction is exothermic; in contrast, negative values of Q indicate absorption of energy so the reaction is endothermic. The corresponding reaction energy levels for 0,1 entrance channel ($E_{01}$) and 2,3 exit channel ($E_{23}$) are shown in Figure 2:

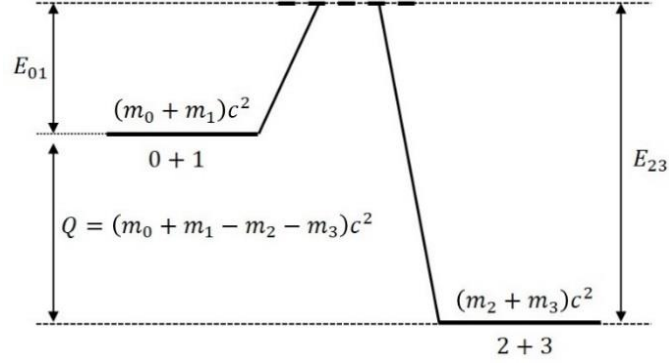

Figure 2 Energy level diagram of nuclear reactions (adapted from Iliadis 2015)

Key points for stellar nuclear reactions to proceed are the exothermic values and the low energy level of the exit channel. If the masses values are known it is possible to calculate the $Q$ values.

## 1.2 Cross Section

In nuclear physics, the event that a nuclear reaction take place is expressed by measuring the probability which represents the rate of one reaction to occur per unit time and volume (Krane 1988). Each nucleus is associated with a geometrical area proportional to the probability of a projectile interacting with that nucleus. This concept is analogous to the probability of shooting a projectile and hitting a target. The target area is called the cross section and is indicated as:

$$\sigma = \pi \left( R_p + R_t \right)^2 \tag{1.4}$$

where $R_p$ and $R_t$ are respectively the radii of the projectile and target. Blatt et al. (1962) have determined that the average nuclear radius R is connected to the atomic number A by:

$$R = R_0 A^{1/3} \tag{1.5}$$

where $R_0$ is a constant value corresponding to $1.3 \times 10^{-13} cm$. Using this relation it is possible for example to calculate the cross-section value for proton-proton reactions. In classical physics, the cross section is only a function of the area but nuclear reactions are also governed by the laws of quantum mechanics, hence the equation can be re-written as:

$$\sigma = \pi \lambdabar^2 \tag{1.6}$$

In this equation λ is the De Broglie wavelength reflecting the wave character of the quantum mechanics process (Blatt et al. 1962) given by:

$$\lambda = \left(\frac{m_p + m_t}{m_t}\right)\frac{\hbar}{\sqrt{2m_p E_l}} \tag{1.7}$$

where $E_l$ is the laboratory energy; $m_p$ and $m_t$ are respectively the mass of projectile and target nucleus and $\hbar$ defined as the reduced Planck. For each nuclear reaction the cross section value varies due the nuclei and the force involved.

## 1.3 Aspects of Quantum Mechanics

Stars on the main sequence including our sun evolve very slow by adjusting the temperature in the core in a way that the average nuclear thermal energy is much smaller compared to the ion-ion Coulomb repulsion value (Clayton 1984, Rolfs 1988, Phillips 1999). This Coulomb barrier is represented in Figure 3:

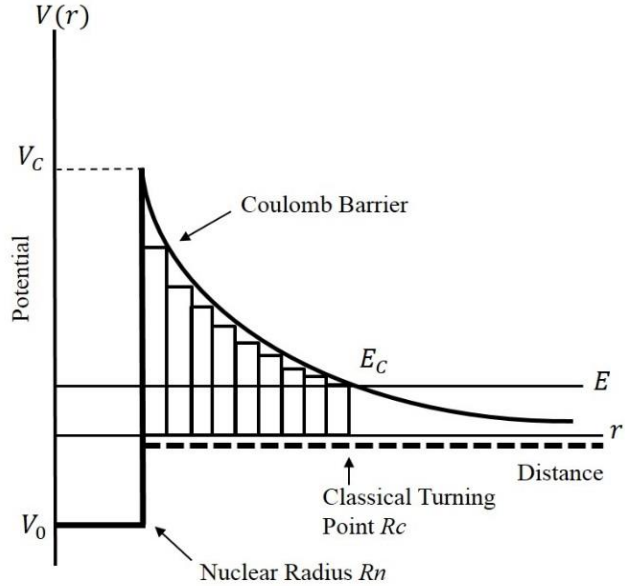

Figure 3 Coulomb barrier representation (adapted from Iliadis 2015)

For classical physics, the Coulomb potential can be written as:

$$V_C = \frac{Z_1 Z_2 e^2}{r} \tag{1.8}$$

and the corresponding height of the Coulomb barrier (see Table 1) related to pp-reaction:

$$E_C = \frac{Z_1 Z_2 e^2}{R_n} = 550\text{keV} \tag{1.9}$$

The classical turning point, defined as the minimum distance at which the nuclei can approach each other with charges $Z_1 Z_2$ and kinetic energy $E_p$ is given by:

$$R_c = \frac{Z_1 Z_2 e^2}{E_p} \qquad (1.10)$$

This means that for the classical physics a pp-reaction can occur only if the kinetic energy is greater than 550keV. Furthermore, since the number of particle is given by the Boltzmann Distribution (see section 1.4 below) only a negligible number of particles can overcome the barrier. Classically a particle cannot penetrate the barrier but quantum mechanically there is a finite value (Rolfs 1988) characterized by the wave function squared ($R_n$ is the nuclear radius), corresponding to the probability of finding the particle at specific position:

$$R_n = |\psi(R_n)|^2 \qquad (1.11)$$

The tunnel effect probability for a particle to penetrate the Coulomb barrier is given by:

$$P = \frac{|\psi(R_n)|^2}{|\psi(R_c)|^2} \qquad (1.12)$$

Bethe (1937) has calculated the transmission probability by solving the Schrodinger equation. Using this method the tunnelling probability $P$ for the pp reaction can be obtained and the corresponding values are shown in table A.1 (Appendix A). In a star's core at low energy regime $E \ll E_C$ hence the classical turning point $(R_c)$ is larger of the nuclear radius and can be approximated to:

$$P \propto exp(-2\pi\eta) \qquad (1.13)$$

where $\eta$ is the Sommerfeld parameter equal to:

$$\eta = \frac{Z_1 Z_2 e^2}{\hbar v} \qquad (1.14)$$

and the exponential term in the equation (1.13) is the Gamow factor that in numerical units correspond to $2\pi\eta = 31.29 Z_1 Z_2 (\mu/E)^{1/2}$ with $E$ given in keV and $\mu$ the reduced mass expressed in amu (Atomic Mass Units):

$$\mu = \frac{m_p m_t}{m_P + m_t} \qquad (1.15)$$

Furthermore, is possible to re-write the cross-section as energy dependence:

$$\sigma(E) = \frac{1}{E} exp(-2\pi\eta) S(E) \qquad (1.16)$$

where $S$ is called the "*Astrophysical S-factor*" (Salpeter 1952) and represents the gradually varying function describing the mass energy inside the stellar core. The dependence of the cross-section and $S(E)$ as function of energy for a reaction ${}^nA(x, y){}^mB$ is shown in Figure 4:

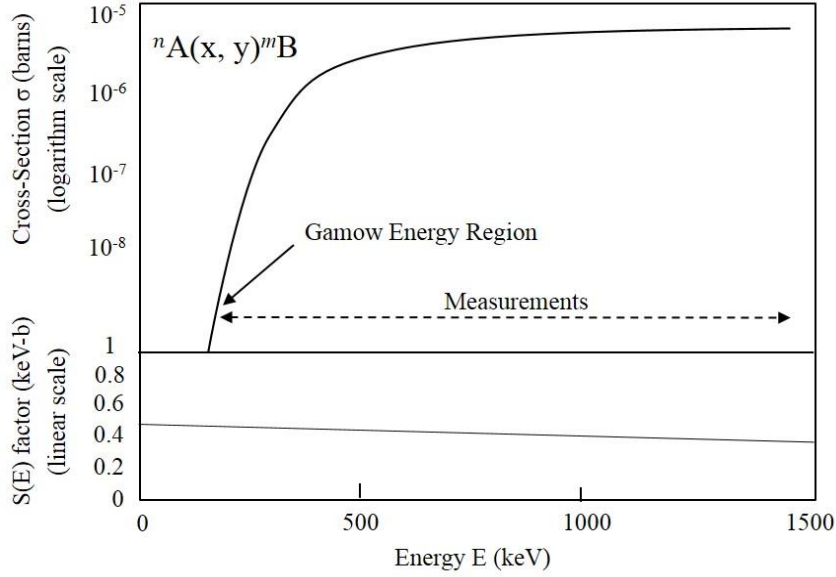

Figure 4 Cross-section and $S$(E) function dependence from the energy (see also Appendix A)

As seen the upper region of the figure (see also A.2 Appendix A) indicates the cross-section for a general *x,y* process on ${}^nA$ to produce ${}^mB$ that drops rapidly as the energy descreases, while the $S$-factor in the lower section remain comparatively constant (Parker 1968, Nagatani 1969, Hilgemeier 1988, Yildiz 2015)

## 1.4 Maxwell Boltzmann Distribution

Particles inside the stars core at temperatute $T$, possesses different velocities and energy which in most cases are connected with the Maxwell Boltzmann distribution:

$$\phi(v) = 4\pi v^2 \left(\frac{\mu}{2\pi kT}\right)^{3/2} exp\left[-\frac{\mu v^2}{2kT}\right] \tag{1.17}$$

In this expression $\phi(v)$ is the velocity probability function. The reaction rates or nuclear cross-section depend also on the relative velocity. Therefore it can be normalised using the energy distribution related to the thermal velocity distribution and the reaction rate per particle pair can be calculated through the following integral:

$$\langle\sigma v\rangle = \int_0^\infty \phi(v)\sigma(v)vdv \tag{1.18}$$

where $\phi(v)$ and $\sigma(v)$ are respectively the velocity probability function and the cross section having velocity in the interval $v$ to $v + dv$. After calculation the final results gives the following expression (for mathematical treatment see appendix A):

$$\langle \sigma v \rangle = \left(\frac{8}{\pi\mu}\right)^{1/2} \frac{1}{(kT)^{3/2}} \int_0^\infty \sigma(E) \, E \exp(-E/kT) dE \qquad (1.19)$$

In this equation, $\sigma(E)$ is the cross section in the stellar environment, $T$ the temperature, $k$ is the Boltzmann constant and $\mu$ is the reduced mass of the interacting particle. Tabulated values of nuclear reaction rates are available in the literature (e.g. Angulo 2010; Dillman 2009; Cyburt 2010). If the astrophysical *S*-factor is included, the thermal reaction rate per pair particle gives:

$$\langle \sigma v \rangle = \left(\frac{8}{\pi\mu}\right)^{1/2} \frac{1}{(kT)^{3/2}} \int_0^\infty S(E) \exp\left[-\frac{E}{kT} - \frac{b}{\sqrt{E}}\right] dE \qquad (1.20)$$

In this equation, $b$ depends on the barrier penetrability while the squared term $b^2 = E_G$ is the Gamow energy given by:

$$E_G = 2\mu(\pi e^2 Z_1 Z_2 / \hbar)^2 = 0.978 \mu Z_1^2 Z_2^2 \, MeV \qquad (1.21)$$

therefore the exponential term for low energy reaction rate is:

$$\exp(-b/\sqrt{E}) = \exp\left(-\sqrt{(E_G/E)}\right) \qquad (1.22)$$

At high energies this term become small thus, the factor with the Maxwell Boltzmann equation $E^{1/2} \exp(-E/kT)$ is negligible. For the thermal averaged reaction rate, a maximum peak value can be calculated by integration of the Maxwell-Boltzmann distribution at $E = kT$ as shown in Figure 5:

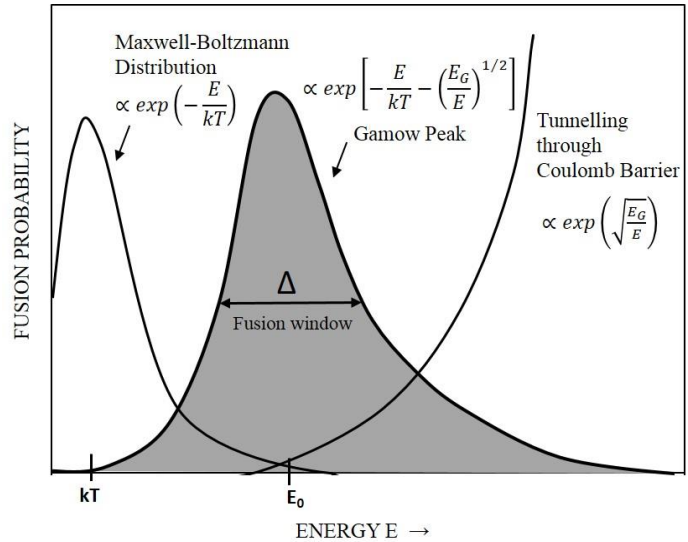

Figure 5 Maxwell-Boltzmann distribution, Gamow peak  and Coulomb barrier transmission probability

When the charge associated with the nucleus increases, the Coulomb barrier gets higher, consequently the Gamow peak (*G.P.*) $E_0$ is shifted to higher energies (Rolfs 1988). The majority of stellar reactions proceed in a narrow band of energies hence the term $S$(E) has almost constant value averaging to $S$(E$_0$). The reaction rate per particles pair can be re-write:

$$\langle \sigma v \rangle = \left(\frac{8}{\pi \mu}\right)^{1/2} \frac{1}{(kT)^{3/2}} S(E_0) \int_0^\infty exp\left(-\frac{E}{kT} - \frac{b}{\sqrt{E}}\right) dE \qquad (1.23)$$

From this integral is possible to calculate different parameters as: $E_0$ (effective burning energy) and $\Delta$ (effective width energy window, Figure 5) related to the thermal averaged reaction at stellar temperature $T_6 = 15$ (where the subscript denotes power of ten $10^x$ so $T_6 = 15$ denotes temperature equal to 15 millions of kelvin). These are summarised and illustrated in Table 1:

Table 1 Thermal reaction rate parameters at $T_6 = 15$

| Reaction | C. B. (MeV) | G. P. (E₀) (keV) | Δ(keV) |
|----------|-------------|------------------|--------|
| $p + p$ | 0.55 | 5.9 | 6.4 |
| $p + N$ | 2.27 | 26.5 | 13.6 |
| $\alpha + {}^{12}C$ | 3.43 | 56 | 19.4 |
| ${}^{16}O + {}^{16}O$ | 14.07 | 237 | 40.4 |

As seen the $E_0$ values are far below the Coulomb barrier (*C.B.*) values at relevant temperature. This means that nuclear reactions burning in stars are so slow because this processes happen at sub Coulomb region allow stars to shine for a long time (Phillips 2010).

## 1.5 Recent Developments

Several worldwide groups have acquired new measurements of the Gamow peak. At Gran Sasso facility new data for ${}^3$He(${}^4$He, 2p)${}^4$He have been collected reducing uncertainty to 10% (Junker et al. 1998). Confortola et al. (2007) have determined a new value of the GamowPeak for ${}^3$He(${}^4$He, γ)${}^7$Be while at Super-Kamiokande centre a new cross-section value has been found for ${}^7$Be(p, γ)${}^8$B which is above the previous result (Junghans et al. 2004). Furthermore, Formicola et al. (2004) have found a new value for the CNO cycle reaction ${}^{14}$N(p, γ)${}^{15}$O corresponding to $S(0) = 1.7 \pm 0.2 keV$ which is 50% lower compared to previous results.

# CHAPTER 2: Hydrogen Burning

## 2.1 The Proton-Proton Chain

The sun has a central temperature of 15.7 million of degrees and shine by burning hydrogen through a process called the pp-chain. The sequence of these reactions is illustrated (Figure 6):

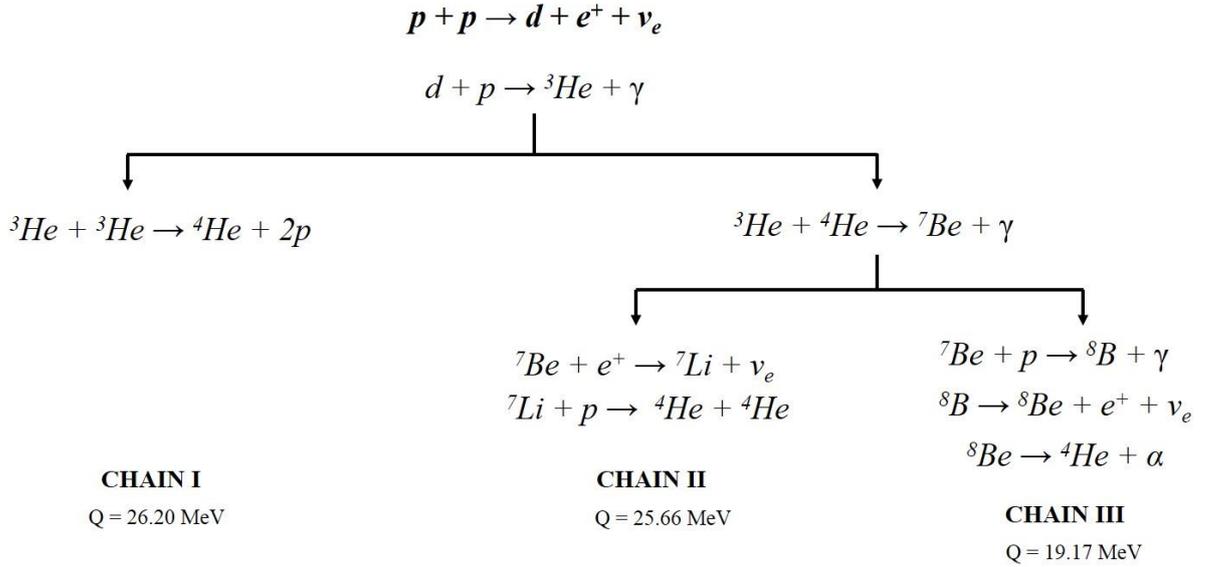

Figure 6 pp-chain reactions scheme

In this reaction cascade the result is the conversion of 4 protons producing 28 MeV, a helium nucleus and the releasing of energy by escaping positrons and neutrinos. The first reaction is:

$$p + p \rightarrow d + e^+ + v_e \qquad (2.1)$$

This proton-proton (pp) reaction (2.1) occurs in two steps: a) penetration of the proton through the Coulomb barrier ($E_C = 550 keV$); b) proton $\beta$-decay process with a positron and neutrino emission. Due to the moderately low Coulomb barrier our Sun would burn the hydrogen fuel in a short time, but the presence of $\beta$-decay processes (2.1) drastically limits the probability for particles as proton or neutron to be captured in the deuterium ($d$) electron valence shell (Bethe and Critchfield 1938). Furthermore, this reaction has low cross-section value below 1MeV hence the surviving time for protons inside the sun is $10^{10}$ years. As consequence of that, a constant radiation is emitted from the sun without explosive events (Bertulani et al. 2016).

In addition, this reaction has been studied by several other groups (Bethe 1938) and its *S-factor* computed (Bahcall 1968, 1969; Kamionkowski 1994; Schiavilla 1998; Butler 2001; Park 2003). After the hydrogen isotope is produced, the deuterium is immediately destroyed through a non-resonant capture-reaction process (see Appendix B) to $^3$He (ground state) as indicated in the following equation (Griffiths 1963, Schmid 1995):

$$d + p \rightarrow {}^3He + \gamma \tag{2.2}$$

The $Q$-value of this reaction corresponds to 5.5 MeV while the *S-factor* at zero energy $S(0) = 2.5 \times 10^{-3} keV$ barn. A schematic capture diagram model is illustrated in Figure 7:

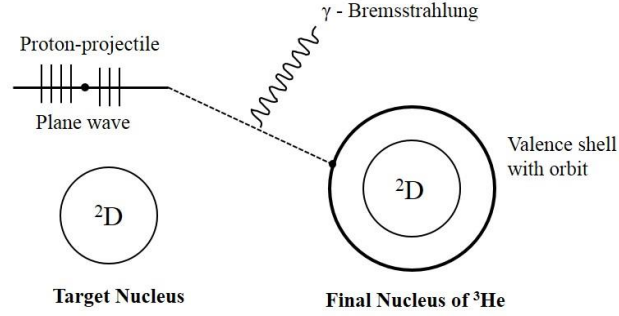

Figure 7 Schematic representation model for the direct capture d(p,γ) $^3$He and photon emission

During that process the proton is absorbed by the $^3$He electronic valence shell orbit and a photon is emitted as bremsstrahlung radiation. Studies have demonstrated that the deuterium/hydrogen ratio in this condition is very small because the deuterium is consumed through the nuclear burning process. Considering $p = H$ and $d = D$ the reaction rates ($r$) for (2.1) and (2.2) are:

$$r_{pp} = \frac{H^2}{2}\langle \sigma v \rangle_{pp} \qquad r_{pd} = HD\langle \sigma v \rangle_{pd} \tag{2.3}$$

Therefore, the evolution of deuterium with time can be calculated by the differential equation:

$$\frac{dD}{dt} = r_{pp} - r_{pd} \tag{2.4}$$

$$\frac{dD}{dt} = \frac{H^2}{2}\langle \sigma v \rangle_{pp} - HD\langle \sigma v \rangle_{pd} \tag{2.5}$$

where $H^2/2\langle \sigma v \rangle_{pp}$ corresponds to the deuterium production rate (2.1) while $HD\langle \sigma v \rangle_{pd}$ is the destruction process (2.2). For a thermodynamic equilibrium and $T_6 = 5$ is possible to write:

$$\left(\frac{D}{H}\right) = \langle \sigma v \rangle_{pp}/\big(2\langle \sigma v \rangle_{pd}\big) = 5.6 \times 10^{-18} \tag{2.6}$$

The observed ratio D/H in the interstellar medium (ISM) is equivalent to $\sim 10^{-5}$ (Liddle 2015). The final equation for the first reaction sequence (pp-chain I) is the following:

$$^3He + {}^3He \rightarrow {}^4He + 2p \tag{2.7}$$

with an *S-factor* value equivalent to 6240 keV barn. However, the amount of deuterium is very low, hence the $^3$He is mainly destroyed by the first reaction despite the similar $S(0)$ *factor* value. For pp-chain II (first step) an electron capture process on $^7$Be atom occur (Brown et al. 2007):

$$^3He + {}^4He \rightarrow {}^7Be + \gamma \tag{2.8}$$

$$^7Be + e^- \rightarrow {}^7Li + \nu_e \tag{2.9}$$

Successively in the second step, the $^7Be$ capturing an electron can leave $^7$Li into the low energetic level (89.6%) or in first excited state (10.4%) as illustrated in Figure 8:

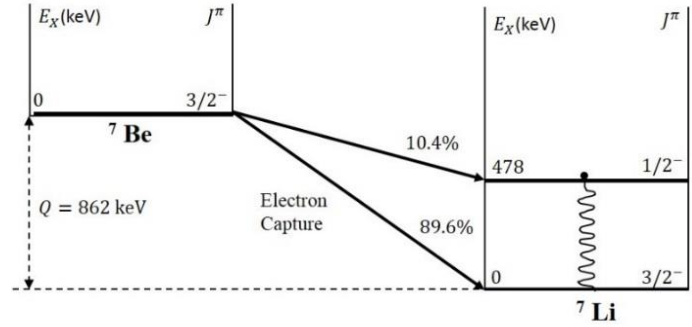

Figure 8 Electron capture on $^7$Be nucleus process with electronic levels and emission of a neutrino

In this process the entire energy corresponds to $Q = 862\ keV$. The electron captured on $^7$Be nucleus can be promoted in the first excited state level of the $^7$Li nucleus ($E_x = 478\ keV$) followed by a decay process to the ground state and emission of a photon (Rolfs 1988). The kinetic energy of this capturing process has been obtained experimentally in laboratory (Rolfs 1988, Iliadis 2015). After the $^7$Be is produced another proton-capture reaction can also occur:

$$^7Be + p \rightarrow {}^8B + \gamma \tag{2.10}$$

In stars not very hot as our Sun this process take place only as 0.02% of the time (Bertulani et al. 2016). The $^8B$ nucleus is radioactive with a lifetime of 1.1s that decay in $^8Be$ nucleus:

$$^8B \rightarrow {}^8Be + e^+ + \nu_e \tag{2.11}$$

This reaction completes the pp chain III with a positron and an electron neutrino emission.

## 2.2 Recent Developments

At the Laboratory for Underground Nuclear Astrophysics (LUNA) new measurements have been obtained for some hydrogen burning reactions (Broggini et al. 2010) such as $^2$H($p,\gamma$)$^3$He, $^3$He($^3$He,2$p$)$^4$He and $^3$He($^4$He,$\gamma$)$^7$Be (See Appendix D). In addition, Junghans et al. (2004) at Super-Kamiokande facility centre have found a cross-section value for the reaction (2.9) of 10% above the previous result (Filippone et al. 1983).

## 2.3 The CN and CNO Cycles

As the temperature increases for stars with $M > 3M_{\odot}$ the CNO cycle starts to dominate. The energy production related to the pp chain and CNO cycle processes as function of the stellar temperature value (expressed as $T_6$) is shown in Figure 9:

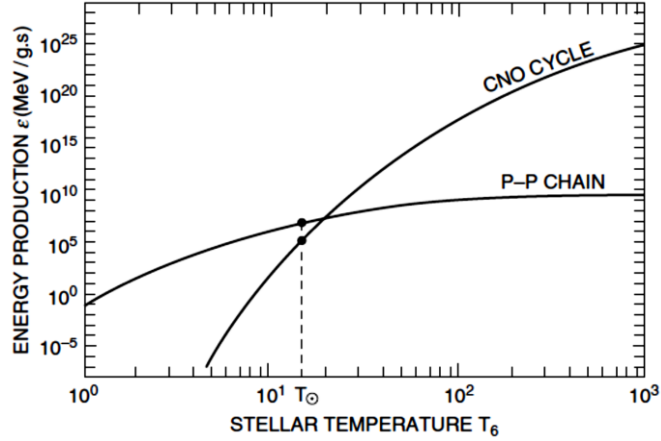

Figure 9 Temperature dependence between pp chain reaction and CNO cycles (Rolfs and Rodney 1988)

As seen at $T_6 \sim 20$ the CNO cycle stars dominates over the pp-chain processes. The CN cycle I sequence discovered by Bethe is part of the CNO Bi-cycle as illustrated in figure 10:

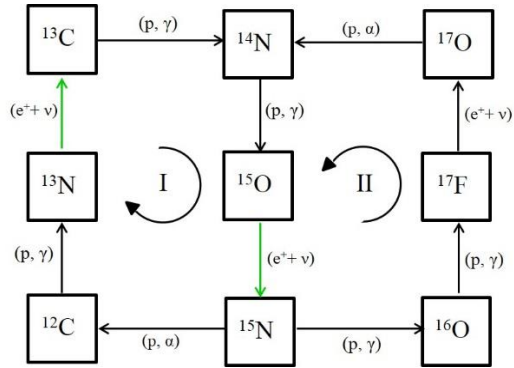

Figure 10 CNO Bi-cycle: cycle I and cycle II processes

In the cycle I two neutrinos are produced during the beta decay of $^{13}N$ and $^{15}O$ processes (green arrows) at relatively low energy since most of the energy generated is retained in the stellar interior (Rolfs 1988). As result of the entire cycle I process 4 protons are converted in helium:

$$4p \rightarrow {}^4He + 2e^+ + 2\nu \qquad (2.12)$$

Since nitrogen isotopes have the highest Coulomb barrier value, electromagnetic force dominates the $^{14}N$ $(p, \gamma)$ $^{15}O$ while nuclear force the $^{15}N$ $(p, \alpha)$ $^{12}C$ process. Therefore, the first reaction is the slowest and determine how quickly the cycle I can proceed. The kinetic energy of this reaction is determined by the half life of the $^{14}N$ nucleus involved in the $^{14}N$ $(p, \gamma)$ $^{15}O$ process $(2.1 \times 10^{12} \, y)$ which is slower by 2 orders of magnitude than other nuclear burning process. The CNO cycle II works at $T_6 > 20$ and starts with $^{15}N$ and $^{16}O$ proton capture process leading $^{17}F$ nucleus that generate $^{17}O$ by beta-decay. This latter *via* proton-alpha process $(p, \alpha)$ yields the $^{14}N$ completing the cycle. However, the cycle II only works in 0.1% of the time compared to the CN cycle I because the *S*-factor for $^{15}N$ $(p, \alpha)$ $^{12}C$ is 1000 larger than for $^{15}N$ $(p, \gamma)$ $^{16}O$ (Bertulani et al. 2016). Very rarely, at higher temperatures new break-out routes can also occur, cycles III and IV (Figure 11):

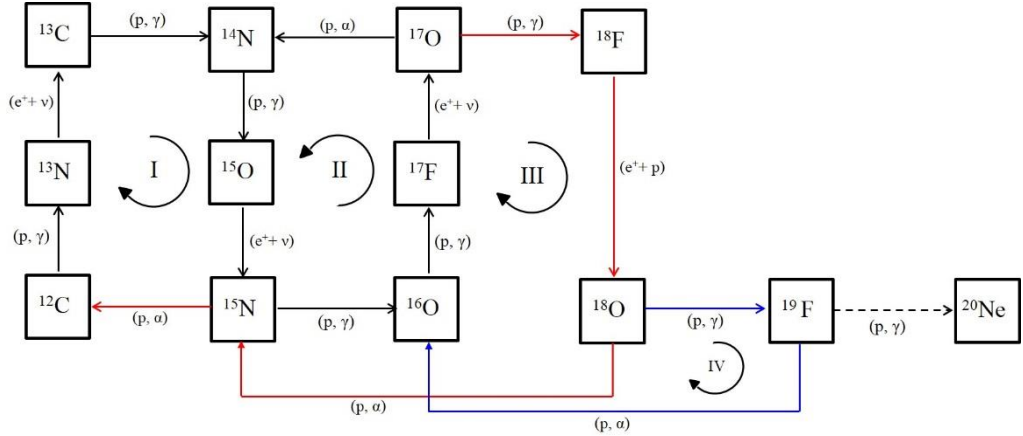

Figure 11 The Four CNO cycles that can occur at high temperatures

The cycles I-IV are also referred as the Four CNO cycles (Bertulani et al. 2016). The cycle III (red arrows) involves the production of $^{18}F$ and $^{18}O$, that after a proton-alpha process $(p, \alpha)$ reaction, returns to $^{15}$N regenerating $^{12}C$ in the cycle I. Furthermore, the $^{18}O$ capturing a proton can evolve in $^{19}F$ and successively by a $(p, \alpha)$ process regenerate $^{16}O$ (cycle IV, blue arrows).

## 2.4 Other Cycles

At high stellar temperatures additional cycles to fuse hydrogen into helium can take place and generate $^{21}$Na or $^{22}$Na from $^{18}$Ne or $^{19}$Ne isotopes as showed in Figure 12:

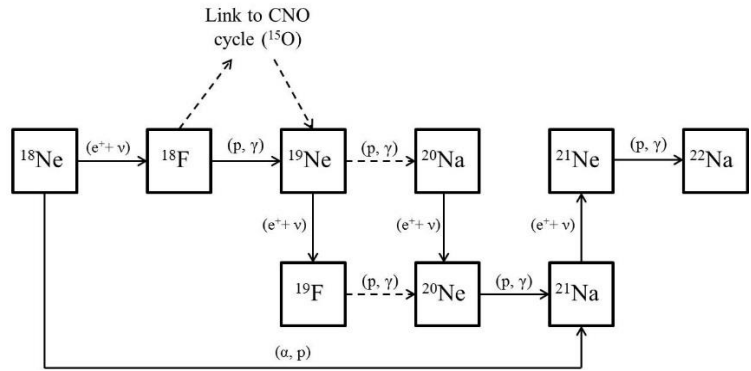

Figure 12 Schematic break-out path formation products from $^{18}$Ne and $^{19}$Ne

The temperature and Coulomb barrier for the $^{18}Ne$ $(\alpha, p)$ $^{21}Na$ process is greater compared to $^{15}O$ $(\alpha, \gamma)$ $^{19}Ne$ reaction. This first nuclear event is also referred as the $(\alpha, p)$ process. However, if the temperature further increases the $^{22}Na$ nucleus can also initiate other break-out routes (Rolfs 1988, Iliadis 2015) as shown in Figure 13:

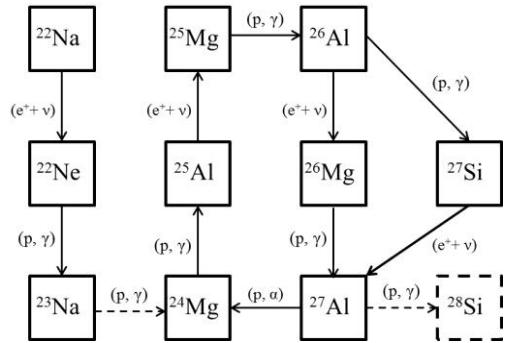

Figure 13 Further possible break-out routes from $^{22}$Na

The NeNa cycle starts with $^{22}Na$ decay process generating a $^{22}Ne$ nucleus with emission of a positron and an electron neutrino. This $^{22}Ne$ nucleus can evolve in $^{23}Na$ through a capture proton process liberating a photon of γ-ray radiation. The $^{23}Na$ has been detected at the surface of some Red Giant stars (Cavallo et al. 1997, Lagarde et al. 2013) while the $^{22}Na$ confirmed in some meteoritic samples (Velbel et al. 2002, Woolfson 2011). Also, the discovery of some inclusions in the Allende meteorite containing an excess of $^{26}Mg$ (Gray et al. 1974, Lee et al. 1976) has suggested the incorporation of radioactive $^{26}Al$ in the material of solar cloud before its condensation phase (Jacobsen et al. 2008, Gaidos et al. 2009, Schiller et al. 2014). The hypothesis of $^{26}Al$ has raised new studies to better understand other possible break-out pathways as the MgAl cycle (right side of Figure 13). In addition, the $^{25}Mg$ $(p, \gamma)$ $^{26}Al$ process can release two isomeric states of $^{26}Al$ nuclei species: $^{26}Al\,^{0}$ and $^{26}Al\,^{m}$ respectively in the ground and excited state. The $^{26}Al\,^{m}$ rapidly decays to $^{26}Mg$ nucleus due the short half-life time ($T_{1/2} = 6.3s$) with emission of an electron neutrino (Figure 14)

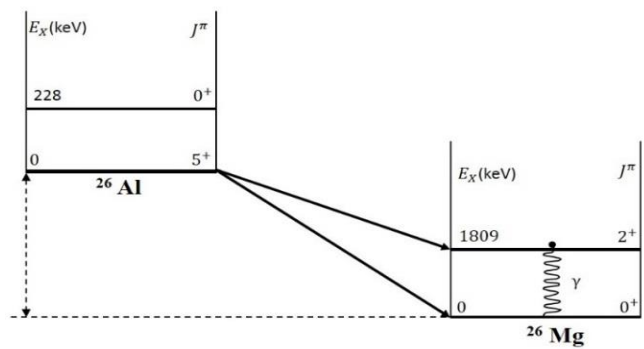

Figure 14 Schematic representation of $^{26}$Al decay process

Due the higher half-life time of $^{26}Al^{\,0}$ in the ground state ($T_{1/2} = 0.72 \times 10^{6} y$) this nucleus can be included during the silicon production via $^{26}Al\ (p,\ \gamma)\ ^{27}Si$ process defining also the amount of $^{26}Al$ produced after MgAl cycle. Furthermore, the $^{27}Al\ (p,\ \gamma)\ ^{28}Si$ provides links to other nuclei break-out paths for the MgAl cycle.

## 2.5 Recent Developments

At the LUNA facility centre the $^{14}N\ (p,\ \gamma)\ ^{15}O$ reaction has been extensively studied (see Appendix D) delivering a rate reduction by a factor of 2 (Constatini et al. 2009; Broggini et al. 2010). This discovery has important astrophysical implications: a) CNO-cycle in the sun contributes 0.8% instead of 1.6% to the total energy production; b) Globular clusters consist mainly of stars with masses smaller than our Sun that burn hydrogen using pp-chain reactions. At the end of their life when the core reaches 0.1 $M_{\odot}$ the nuclear energy generated during the H-burning become inadequate hence, the stellar core contracts, the temperature rises and the CNO-cycle take place over the pp-chain reaction process. Thus, the migration of these stars from the main sequence is directly linked to the CNO-cycle rate and consequently to the $^{14}N\ (p,\ \gamma)\ ^{15}O$ reaction. As a consequence of the new LUNA measurement reduction values, the age of Globular clusters increase by 1 billion (Imbriani et al. 2004) to 14 billion years which is more compatible with the estimate age of the universe; c) resolution of the discrepancy about the over-production of chemical elements heavier than carbon between observation data and stellar models in AGB stars (Herwig 2006).

# CHAPTER 3: Helium and Heavier Elements Burning

## 3.1 Triple-α Capture Process

After the hydrogen is consumed ($7 \times 10^6$ years), the core slowly contracts and its temperature increases releasing kinetic energy as result of the gravitational energy transformation (Ryan and Norton 2002). During this contraction time, the residual hydrogen ignites the edge of the helium core. This process of helium burning and conversion into $^{12}C$ occurs in two steps:

$$\alpha + \alpha \rightarrow {}^8Be \tag{3.1}$$

$$\alpha + {}^8Be \rightarrow {}^{12}C \tag{3.2}$$

In the first step two helium nuclei ($\alpha = {}^4He$) react together giving $^8Be$. The relative equilibrium of $^8Be$ can be estimated using the Saha equation:

$$N_{12} = \frac{N_1 N_2}{2} \left( \frac{2\pi}{\mu kT} \right)^{3/2} \hbar^3 \omega \, exp\left( -\frac{E_R}{kT} \right) \tag{3.3}$$

The Saha equation defines the relation between free particles and those bound in atoms assuming ideal gas conditions and neglecting the ion-ion interactions. In the equation (3.3) $N_1$ and $N_2$ corresponds to the number density respectively for 1 and 2 particle, $\mu$ is the reduced mass, $E_R$ is the resonance energy and $\omega$ is the statistical factor related to the spin state $((2J + 1))/(2J_1 + 1)(2J_2 + 1)$ for the nuclei 1 and 2 respectively indicated as $J_1$ and $J_2$ (see also Appendix B and C). At temperature $T_6 = 11$ and density $\rho = 10^5 g/cm^3$ the ratio of $^8Be/^4He$ is $\sim 10^{-10}$ (Bertulani et al. 2016). This low amount of $^8Be$ seems to play a key role (Salpeter et al. 1954) in the triple α-capture process, equation (3.2). A schematic representation of these reactions (3.1) and (3.2) with energy levels is shown in Figure 15:

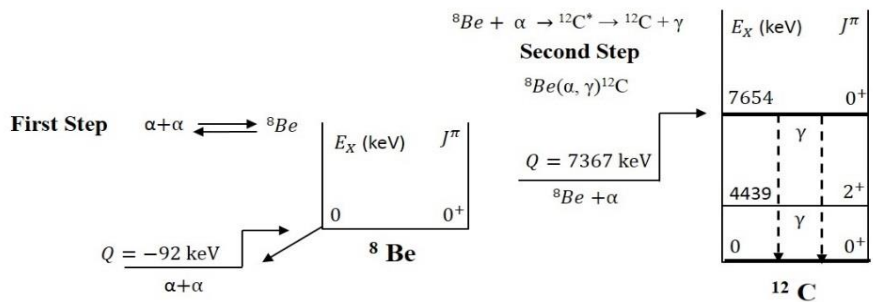

Figure 15 Triple alpha processes for $^{12}$C synthesis showing the two steps

The $^{12}C$ formation proceed through an excited state (resonance) where the excited carbon nucleus can be indicated as $^{12}C^*$ (Hoyle 1954). The details of this process are described later in this section. To calculate the reaction rate for $^{12}C$ formation is possible to use the resonant state (Appendix C) and the thermally averaged cross-section equation (Rolfs and Roodney 1988):

$$r_{3\alpha} = N_{8_{Be}} N_\alpha \langle \sigma v \rangle_{8_{be}+\alpha} \tag{3.4}$$

In this expression $N_{8_{Be}}$ and $N_\alpha$ are respectively the density numbers of $^8Be$ and $^4He$ nuclei while the term $\langle \sigma v \rangle$ is related to the Maxwell Boltzmann distribution $\psi(E)$ for thermal average (see Chapter 1). The reaction rate integral for the equation (3.4) can be written (Clayton 1984):

$$r_{3\alpha} = N_{8_{Be}} N_\alpha \int_0^\infty \psi(E) v(E)\, \sigma(E) dE \tag{3.5}$$

where $\psi(E)$ is the Maxwellian term equivalent to (Clayton 1984):

$$\psi(E)dE = \frac{2}{\sqrt{\pi}} \frac{E}{kT} exp\left(-\frac{E}{kT}\right) \frac{dE}{\sqrt{kTE}} \tag{3.6}$$

and $\sigma(E)$ is the resonance strength (Appendix C) that can be represented including the total number of decay channel widths (Iliadis 2015):

$$\sigma(E) = \pi \lambda^2 \frac{(2J+1)}{(2J_1+1)(2J_2+1)} \frac{\Gamma_1 \cdot \Gamma_2}{(E-E_R)^2+(\Gamma/2)^2} \tag{3.7}$$

In this equation $\pi \lambda^2$ is the geometrical factor proportional to 1/E; $((2J+1))/(2J_1+1)(2J_2+1)$ is the spin factor ($J$ = spin state of the nuclei, $J_1$ = spin of the projectile; $J_2$ = spin of the target) and $\Gamma_1 \cdot \Gamma_2/(E-E_R)^2 + (\Gamma/2)^2$ is a strongly energy dependent term ($\Gamma_1$ = probability of compound formation via entrance channel; $\Gamma_2$ = probability of compound decay via exit channel; $\Gamma$ = total width of compound's excited state; $E_R$ = resonance energy). Furthermore, considering that $^{12}C^*$ can also broken into three $\alpha$-particles ($3\alpha$), $\Gamma_1 = \Gamma_\alpha$ dominates over $\Gamma_2 = \Gamma_\gamma$ hence $(\Gamma_1 \Gamma_2/\Gamma) \sim \Gamma_2$ and the concentration of $^8Be$ (at the equilibrium) inside the $\alpha$ particle nuclei environment can be found using the Saha's equation (Clayton 1984):

$$N_{8_{Be}} = N_\alpha^2 \omega f \frac{h^3}{(2\pi \mu kT)^{3/2}} exp\left(-\frac{E_R}{kT}\right) \tag{3.8}$$

The term $f$ is the screening factor that includes the cross section increasing value caused by the electron shield in the stellar environment. In stars atoms are fully ionised, but the electron gas still shield the nucleus increasing the Coulomb Barrier penetrability and cross-section value. Therefore, the triple $\alpha$-reaction rate can be calculated using the $^{12}C^*$ excited state as indicated:

$$^8Be + \alpha \rightarrow {}^{12}C^* \tag{3.9}$$

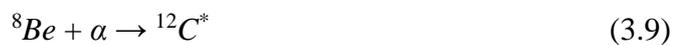

The resulting $^{12}C^*$ concentration value can be expressed as stellar reaction rate for a narrow resonance per particle pair (See Appendix C for mathematical treatment):

$$r_{3\alpha} = N_{8_{Be}}N_\alpha\hbar^2\left(\frac{2\pi}{\mu kT}\right)^{3/2}(\omega\gamma)fexp\left(-\frac{E_R'}{kT}\right) \tag{3.10}$$

where $\mu$ corresponds to the reduced mass of $^8$Be and $\alpha$-particle. Combining the equations (3.8) and (3.10) with $(\omega\gamma) = \Gamma_\alpha\Gamma_\gamma/\Gamma$ is possible to write (Rolfs and Rodney 1988):

$$r_{3\alpha\to^{12}C} = \frac{N_\alpha^3}{2}3^{3/2}\left(\frac{2\pi\hbar^2}{M_\alpha kT}\right)^3 f\frac{\Gamma_\alpha\Gamma_\gamma}{\Gamma\hbar}exp\left(-\frac{Q}{kT}\right) \tag{3.11}$$

In this equation $\Gamma_\alpha$ is the $\alpha$-particle width, $\Gamma_\gamma$ is the total electromagnetic decay width of $^{12}C$ in the ground state and $Q$ corresponds to $E_R(8_{Be} + \alpha) = 287\ keV$ while $E_R(\alpha + \alpha) = |Q| = 92\ keV$ (Figure 15). The energy product rate is given by the equation (Clayton 1984):

$$\epsilon_{3\alpha} = \frac{r_{3\alpha}Q_{3\alpha}}{\rho} \propto \frac{\rho^2 X_\alpha^3}{T_8^3}fexp(k/T_8)\ erg\ gm^{-1}s^{-1} \tag{3.12}$$

The triple alpha process is more efficient at high temperatures value e.g. $T_8 = 1$. This suggest that the helium at this temperature and density is highly reactive because the *He* shell is suddenly activated (Clayton 1984; Rolfs and Rodney 1988). This phenomenon in the stellar core is also called *helium flash* that can reach a luminosity value above $10^{11}L_\odot$.

## 3.2 Asymptotic Giant Branch (AGB) Stars

As a consequence of hydrogen exhaustion and increasing temperature, the outer layers expand and cool down increasing the star's luminosity to become a *red giant*. In this process the stars move on the right side (Red Giant branch), successively on the left side (Horizontal Branch) of the Hertzsprung-Russell (HR) diagram as shown in Figure 16:

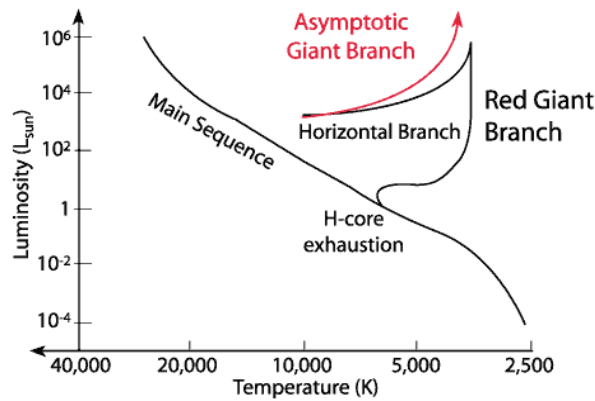

Figure 16 Hertzsprung-Russell diagram with Asymptotic Giant Branch (AGB) path in red

Finally the stars move again on the upper-right region in the *Asymptotic Giant Branch* (AGB) of the HR diagram. AGB stars emits into the interstellar medium new synthesized material in form of stellar winds causing the exhaustion of its envelope in million of years.

## 3.3 Carbon Burning

In AGB stars reaction includes the burning of $^{12}C$ by a triple alpha capture process into $^{16}O$ as:

$$^{12}C + \alpha \rightarrow {}^{16}O + \gamma \tag{3.13}$$

This reaction is very complex, involving resonance capture, non-resonant capture and sub-threshold resonance processes (see Appendix B and C) having an *S*-factor value corresponded to 0.3 MeV (Bertulani et al. 2016). In order to calculate the reaction rate the Saha equation as seen previously for the triple α-process can be used (Bertulani et al. 2016):

$$r_{12_C} = \left(\frac{n_\alpha}{7.5 \times 10^{26}/cm^3}\right)(2.2 \times 10^{13}/s)e^{-69/T_8^{1/3}}T_8^{-2/3} \tag{3.14}$$

This equation is highly temperature dependent. It is also important to note that red giants are the main galactic source of $^{12}C$ and $^{16}O$. The energy diagram levels for $^{16}O$ for a temperature value of ~ $T_9 = 2$ are illustrared in Figure 17:

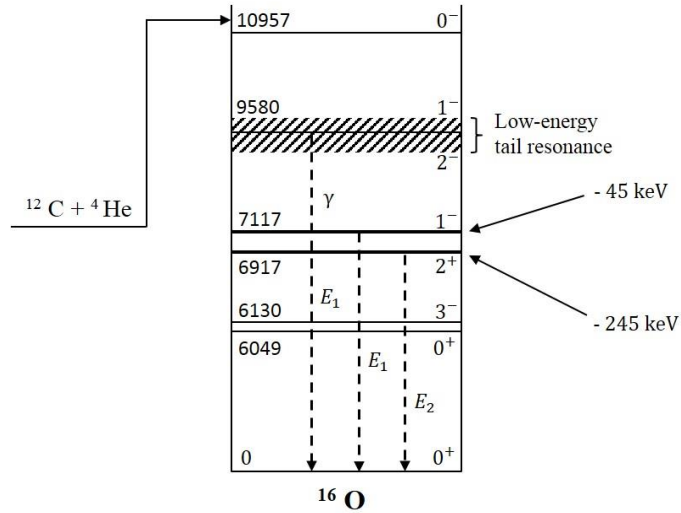

Figure 17 Energy levels of $^{16}O$ nucleus with $^{12}C$ alpha particle capture

The $^{12}C(\alpha, \gamma)^{16}O$ reaction rate depends on the low energy tail of the broad resonance energy value located at 9580 keV (400 keV width) and in addition by two sub-level energy lines respectively at $E_R = -45$ keV and $E_R = -245$ keV (Rolfs and Rodney 1988). The emission of a photon (dashed-lines) carries away several rotational modes and components of the electric field making complicated the cross-section measurement value.

For stars having mass $M > 8 - 10M_\odot$ the core further contract and as the temperature reach high values the carbon nuclei start to react each other. The fusion of two $^{12}C$ produces a $^{24}Mg$ nucleus having an excited energy of 14 MeV. The Gamow energy corresponds to 1 MeV and the reaction rate of $^{12}C$ (per pair) can be found using the equation (Clayton 1984):

$$\log r_{12,12} = \log f_{12,12} + c_1 - \frac{c_2(1 + 0.1T_9)^{1/3}}{T_9^{1/3}} - \frac{2}{3}log(T_9) \tag{3.15}$$

where $f$ is the electron screening factor while $c_1$ and $c_2$ are two numbers. During the $^{12}C$ burning the produced proton and α-particles are rapidly consumed by the sequence:

$$^{12}C\,(p, \gamma)^{13}N(e^+, \nu_e)^{13}C(\alpha, n)^{16}O \tag{3.16}$$

As discussed in Chapter 2 the reaction $^{12}C\,(p, \gamma)^{13}N$ is also involved in the CNO cycle and in the equation (3.16) where the proton is transformed in a neutron releasing a α-particle during the conversion from $^{12}C$ to $^{16}O$ nuclei. However in some AGB stars these protons can penetrate across the helium-shell and generate also $^{13}C$ which justify the presence of this element inside that class of stars. Carbon burning process can also generate other species as Ne, Na and Mg by the following reaction sequences (Rolfs and Rodney 1988):

$$^{12}C + {}^{12}C \rightarrow {}^{20}Ne + {}^{4}He \tag{3.17}$$

$$^{12}C + {}^{12}C \rightarrow {}^{23}Na + p \tag{3.18}$$

$$^{12}C + {}^{12}C \rightarrow {}^{23}Mg + n \tag{3.19}$$

Due to the higher Coulomb barrier reactions like $^{12}C + {}^{16}O$ and $^{12}C + {}^{20}Ne$ proceed with lower reaction rate values compared to $^{12}C + {}^{12}C$ process (Bertulani et al. 2016).

## 3.4 Neon and Oxygen Burning

After the carbon is exhausted further core contraction cause the temperature to raise again and at $T_9 \sim 1$ energetically rich photons start to be absorbed through a process involving the disintegration of $^{20}Ne$:

$$^{20}Ne + \gamma \rightarrow {}^{16}O + {}^{4}He \tag{3.20}$$

At this temperature the most common photodisintegration reactions for $^{20}Ne$ are $(\gamma, n), (\gamma, p), (\gamma, \alpha)$ processes (Bertulani et al. 2016). The photonuclear decay rate is:

$$r(\gamma, \alpha) = \left[exp\left(-\frac{E_x}{kT}\right)\frac{2J_R + 1}{2J_0 + 1}\frac{\Gamma_\gamma}{\Gamma}\right]\frac{\Gamma_\alpha}{\hbar} \tag{3.21}$$

where the squared backet term refers to the probability to find the nucleus in the excited state $E_x$ having spin $J_R$ while the other term ($\Gamma_\alpha/\hbar$) represents the decay rate of the excited state with the emission of α-particle (Rolfs and Rodney 1988). Considering that $E_X = E_R + Q$ it gives:

$$r(\gamma,\alpha) = \frac{exp(-Q/kT)}{\hbar(2J_0 + 1)}(2J_R + 1)\frac{\Gamma_\alpha\Gamma_\gamma}{\Gamma}exp\left(\frac{E_R}{kT}\right) \qquad (3.22)$$

The process of neon burning at the end leaves a mixture of α-particles, oxygen $^{16}O$ and magnesium $^{24}Mg$. During oxygen burning the core contracts again and the temperature increases reaching $T_9 \sim 2$ where $^{16}$O nuclei can start to react together giving:

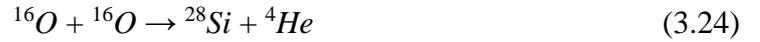

$$^{16}O + {}^{16}O \rightarrow {}^{28}Si + {}^4He \qquad (3.24)$$

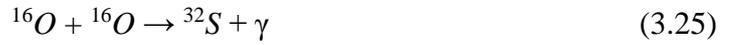

$$^{16}O + {}^{16}O \rightarrow {}^{32}S + \gamma \qquad (3.25)$$

These reactions occur in 50% of the time rate and have a Q-value of 9.593 MeV. The $^{28}Si$ nuclei at $T_9 \sim 3$ start to burn but since the Coulomb barrier is too high they can't react alone. This process instead occur by a sequence of alpha particles, neutron and protons additions and will be discussed in the Chapter 4. A summary of the nuclear burning conditions and time for a 25 $M_\odot$ solar mass star are represented in table 2 (Phillips 2010):

Table 2

| Burning | Time (years) | Temperature ($10^9K$) | Density $g/cm^3$ |
|---|---|---|---|
| *Hydrogen* | $7 \times 10^6$ | 0.06 | 5 |
| *Helium* | $5 \times 10^5$ | 0.2 | $7 \times 10^2$ |
| *Carbon* | 600 | 0.9 | $2 \times 10^5$ |
| *Neon* | 1 | 1.7 | $4 \times 10^6$ |
| *Oxygen* | 0.5 | 2.3 | $1 \times 10^7$ |
| *Silicon* | 1 day | 4.1 | $3 \times 10^7$ |

It is also important to highlight that hydrogen and helium burning occur for stars having mass $0.8\ M_\odot - 8\ M_\odot$ while carbon to silicon burning occur during the later stages of stars with masses larger than $8\ M_\odot$

# CHAPTER 4: Explosive Burning

## 4.1 Silicon Burning

When the oxygen fuel is exhausted the core contains maily $^{28}$Si. At this stage the core further contracts increasing density and temperature ($T_9 > 3.5$). Due the high Coulomb barrier silicon $^{28}$Si atoms cannot react with itself ($^{28}$Si + $^{28}$Si) since they are destroyed by photodisintegration before they can reach the necessary temperature. In these conditions silicon atoms are immersed inside an high concentration of high-energy γ-rays. The silicon burning process is highly complicated and has been extensively studied (Clayton and Flower 1968; Woosley, et al. 1973; Iliadis 2015). The process starts with a series of photodisintegration reactions where $^{28}$Si nuclei are transformed by gamma-ray protons into $^4$He nuclei and a mixture of free protons and neutrons. Since not all $^{28}$Si are destroyed, the alpha-particles are captured by the undissociated $^{28}$Si nuclei giving higher mass nuclei (Ryan and Norton 2010; Iliadis 2015) as shown in Figure 18:

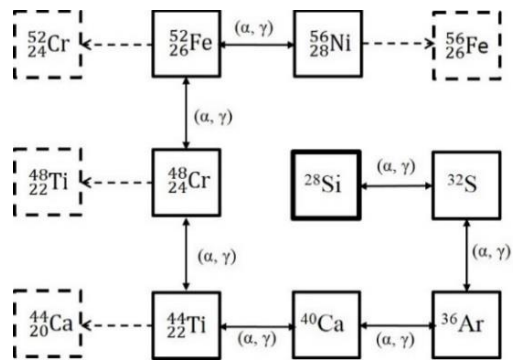

Figure 18 Simplified Silicon Burning process with stable decay nuclei in black dashed-lines

In this diagram the $^{28}$Si nuclides (bold-black square, Figure 18) are converted into iron peak nuclides such as $^{56}_{26}$Fe. A series of reactions to produce more massive nuclei take place giving different nuclei as sulfur, argon, calcium, titanium, chromium and iron. All these processes are in equilibrium therefore are indicated with double head arrows. This sequence of reactions (also called *Silicon burning*) allows stars to generate a further amount of energy (Rolfs and Rodney 1988). The nuclei indicated as dashed-line ($^{44}_{20}$Ca, $^{48}_{22}$Ti, $^{52}_{24}$Cr and $^{56}_{26}$Fe) are more stable and result from decay processes of unstable isotopes as $^{44}_{22}$Ti, $^{48}_{24}$Cr, $^{52}_{26}$Fe and $^{56}_{28}$Ni having an half-lives of order of a few hours to years (Woosley et al. 1973). Due to the number of photodissociation reactions involved, where the nuclei are broken down, this process is also referred to as *Silicon melting* where the silicon nuclei are immersed in a sea of helium nuclei (Rayn and Norton 2010). The Typical time-scale for the entire process is only 1 day. A schematic diagram showing photodisintegration (with excited state $E_x$) is shown in Figure 19:

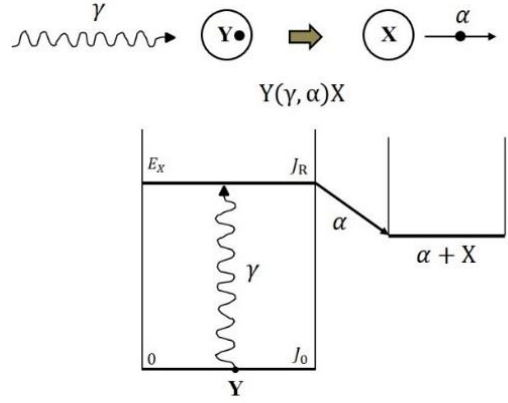

Figure 19 Photodisintegration process for a general Y(γ,α)X reaction

A Y nucleus in the ground state is promoted to the excited state by gamma radiation and broken down into X nucleus releasing α-particles. In this process $J_0$ and $J_R$ are respectively the spins of the ground and excited state that can be described by the Boltzmann probability function:

$$P(E_x) = exp\left(-\frac{E_x}{kT}\right)\frac{2J_R + 1}{2J_0 + 1}$$

This equation can be re-written as:

$$P(E_x) = exp\left(-\frac{E_x}{kT}\right)\frac{2J_R + 1}{2J_0 + 1}\frac{\Gamma_\gamma}{\Gamma}$$

where the factor $\Gamma_\gamma/\Gamma$ includes the fraction of nuclei in the excited state that can decay with emission of α-particles.

## 4.2 Beyond silicon burning: Supernova explosion

After the final burning process of silicon nuclei inside the core of a massive star $\left(M \geq 8M_\odot\right)$ a multilayer structure similar to an onion take place. These layers have different composition divided by "hydrostatic nuclear burning shells" wrapping the central core made predominantly of iron $^{56}$Fe nuclei. Stars with different masses are characterised by different structures, shell composition and location of the nucleosynthesis processes. A star having initial solar composition and a mass equivalent to $25M_\odot$ contains a central core maily composed of $^{56}$Fe as result of the silicon burning events (Iliadis 2015). The next/first shell outside the core contains $^{28}$Si nuclei produced by the oxygen layer burning process. During this time, the conversion process $^{28}Si \rightarrow {}^{56}Fe$ remain still active at the boundary between the silicon shell and the iron core; in contrast the reaction $^{16}O \rightarrow {}^{28}Si$ persists between the oxygen and silicon shell region. The onion-structure is illustrated in Figure 20:

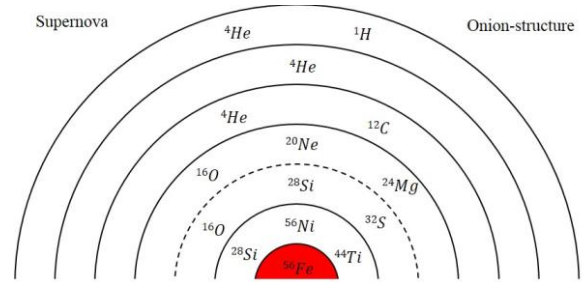

Figure 20  Evolution and Structure of massive star M>25 predicted by one and spherical dimensional models

After this point heavy elements are ignited producing a nuclear burning process localised between the different overlaid shells. As result of this process the core's mass further increases. When the core mass reaches the Chandrasekhar limit $\left(M \sim 1.4 M_{\odot}\right)$ the electron degeneracy pressure cannot balance gravity and consequently the core start to collapse (Jordi et al. 2011, Iliadis 2015). This process is very complex but it can be summarised in three main steps: *Electron Capture*; *Neutrino Trapping* and *Conversion to Nuclear Matter* (Figure 21).

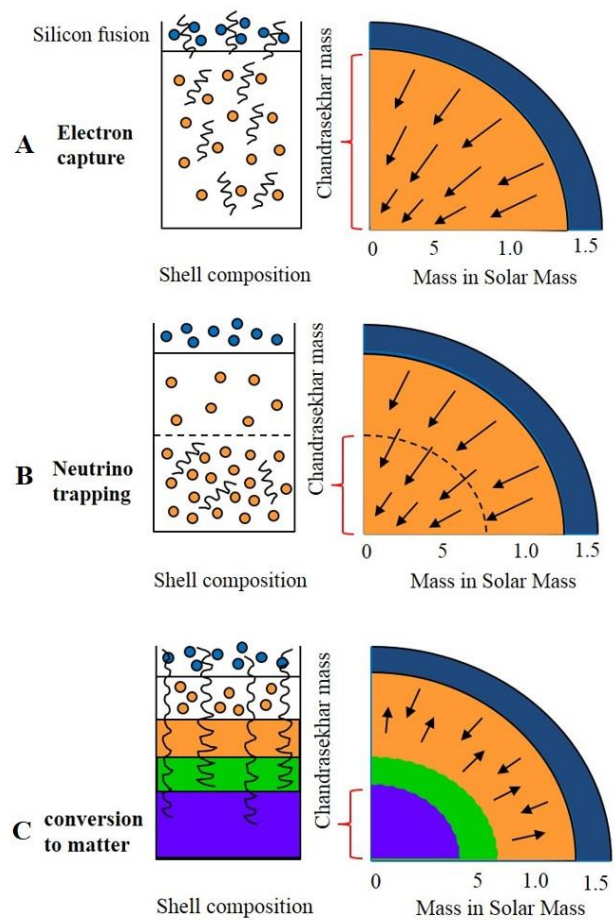

Figure 21 Supernova core-collapse steps: nuclear matter (violet), heavy nuclei (green), iron (orange), silicon (blue)

*Electron Capture* (Figure 21A). In this stage the electron capture process accelerates favouring the inward movement towards the core, converting a proton ($p$) and electron ($e$) into a neutron ($n$) and neutrino ($v$) as shown in the following equation:

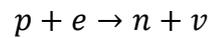

$$p + e \rightarrow n + v$$

This process of electron conversion reduce the electron pressure and consequently the Chandrasekhar mass (Rolfs and Rodney 1988). This latter is a function of the electron and nuclei number: an high value of electrons correspond to high value of electron pressure hence to a large Chandrasekhar mass.

*Neutrino Trapping* (Figure 21B). When the density of the core reach a value of $4 \times 10^{11} g/cm^3$ there is significant transformation of the physical conditions and the matter become opaque to neutrinos (Iliadis 2015). In this scenario, neutrinos (short black corrugated lines) are trapped inside the core. Neutrinos are particle species that interact very weekly with any other form of matter making neutrino detection on Earth very tricky to achieve. When the core reaches high density the neutrinos are packed very tight and they cannot escape. At this stage the entire Chandrasekhar mass ($< 1M_{\odot}$) collapse and the central region is converted into nuclear matter.

*Conversion into nuclear matter* (Figure 21C). During this step the produced nuclear matter is compressed over the equilibrium density and rebounds emitting powerful shock waves (long corrugated black lines). Nuclear matter is a hypothetical system composed of nucleons (protons and neutrons) distributed at some uniform density over the volume. Nuclear matter is incompressible and can reach a density value of $2.7 \times 10^{14} g/cm^3$. Neutron stars/Pulsars and Supernovae remnants are the best candidates for a physical representation of this physical status. The conversion in nuclear matter is the trigger for supernovae explosion followed by the sudden release of gravitational energy (Iliadis 2015).

## 4.4 Simulation Models

In recent decades considerable efforts have been made about core-collapse mechanisms of supernovae but despite remarkable progresses the real processes beyond these events remain still unknown. Observational data of neutrinos produced during the SN 1987A explosion have established only some fundamental principles of the supernovae core-collapse mechanism (Bionta et al. 1987; Hirata et al. 1987; Prahan and Sultana 2011). The SN 1987A supernova remnant as composite image is illustrated in Figure 22:

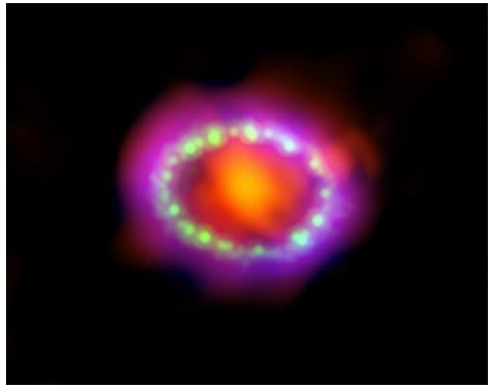

Figure 22 SN 1987A supernova remnant composite image from X-Chandra Observatory and Hubble Space Telescope

A pioneering work carried out by Wilson and Bethe (1985) used the *delayed-neutrino explosion mechanism* where neutrinos are able to increase the temperature around the shock area and regenerate new shock events in a time frame of milliseconds delay. One dimensional simulations (1D) applying *the neutrino-transport approach* have failed to validate supernova explosions due to numerical challenges that involve nuclear physics, general relativity, and neutrino flavours (Rampp and Janka 2002; Buras et al. 2003). In this model, the thermal energy is carried away from the core by neutrinos. These particles are also referred to as radiating neutrinos and are located with antineutrinos inside the neutrinosphere (Figure 23):

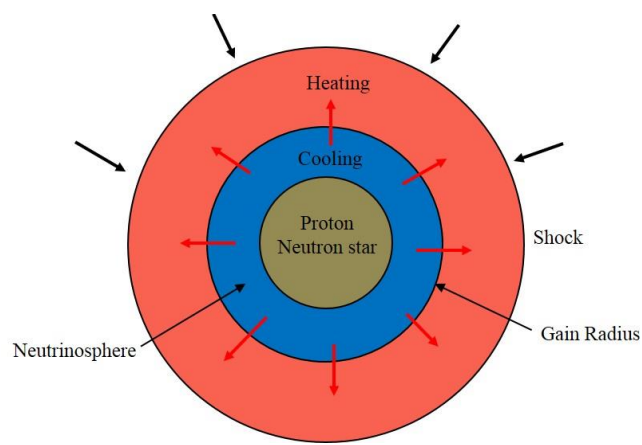

Figure 23 Schematic representation of neutrino-transport mechanism (spherical symmetry model)

However, 1D simulations involving a spherical symmetry has been successfully to validate supernova explosions in the situation where stars contain oxygen, neon, magnesium and a low mass iron-core (Kitaura et al. 2006; Janka et al. 2008; Fischer et al. 2010). Asymmetric two dimensional simulations (2D) have delivered explosion events below the characteristic energy value $(10^{51} \text{erg})$ necessary for the core-collapse (Janka et al. 2012; Muller et al. 2014; Nakamura et al. 2015). Despite these studies, recently Muller (2017) has demonstrated that asymmetries are essential for the explosion mechanism. This three dimensional simulations (3D) involving *the neutrino-driven approach* is shown in Figure 24:

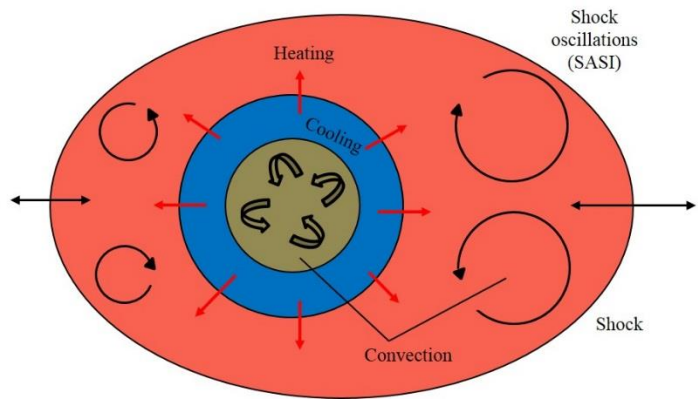

Figure 24 Schematic representation of neutrino-driven explosion mechanism (adapted from Muller 2017)

In this model the core emits neutrinos through the blue layer inside a proto-neutron star (PNS) region (red), but a small fraction are reabsorbed around the increasing area after the shock event (Muller 2017). Convection processes are generated in this region as result of the increasing temperature from the neutrino heating process. Furthermore the Standing Accretion Shock Instability (SASI) can also produce giant oscillation shock waves (Blondin et al. 2003). However, for massive shock waves remarkable improvement have been made in 3D simulations using the critical limiting luminosity $(L_{v,crit})$ related to the mass accretion rate $(\dot{M})$ considering a static spherical symmetry (Summa et al. 2016). Recently, the same team has achieved a successful 3D and multi-D explosion simulations introducing sophisticated analytical equations that include observable variables (*i.e.* critical luminosity, mass accretion rate, gain radius) and other theoretical terms. For multi-D simulations they have also introduced the turbulent flow term previously reported by Muller and co-workers (2015). This term describes the thermodynamic, pressure and flow-velocity evolution related to the outgoing gas. Although relevant progresses have been made in Supernoavae explosion simulations a rigorous mathematical theory remain extremely complex to develop due the nature of the processes involved.

## Conclusions

This dissertation has focused on astrophysical nuclear reactions involving in low mass stars (i.e. our Sun) and massive stars. Considerable progresses have been made in this field due to advances in technologies that use sophisticated satellites as astronomical tool observatories. Nuclear physics, cross section, Coloumb Barrier and the Maxwell-Boltzmann distribution including recent development in the field have been discussed. Proton-proton chain reactions, cold and hot CNO cycles together with other cycles have been studied including recent measurements at LUNA facility centre. This latter have revealed important results data consequences about age of globular clusters, contribution to CNO cycle and age of the universe. The kinetic of triple-alpha reactions, carbon, neon oxygen burning have been also studied. A short overview of the silicon burning process has revelead a complicated nuclear path and photodisintegration processes involved. In addition some supernovae simulation models have explored the dynamics behaviour of a number of mathematical models mainly in 3D. However many problems remain unsolved due the limited capability to reproduce these reactions with current technologies in terrestrial collider facilities. In the near future the building of new high technological nuclear physics centre will play a key role to connect the microscopic dynamics of nuclear systems and some big stellar questions (i.e. dark matter, big-bang, stellar formation, nucleosynthesis and supernova explosions). Deeper understanding of the nuclear structure and nuclear reactions will be crucial to reveal the mechanism of core-collapse supernovae explosions.

# Appendix A

## A.1 Tunnelling probability for pp reaction as function of Centre-of-Mass Energy $E$

| Tunnelling Probability | | |
|---|---|---|
| **Energy (keV)** | **Experimetal result** | **Simulated** |
| 1 | $8.9 \times 10^{-10}$ | $2.5 \times 10^{-10}$ |
| 2 | $5.6 \times 10^{-7}$ | $1.6 \times 10^{-7}$ |
| 5 | $1.7 \times 10^{-4}$ | $5.0 \times 10^{-5}$ |
| 10 | $3.1 \times 10^{-3}$ | $9.2 \times 10^{-4}$ |
| 20 | $2.4 \times 10^{-2}$ | $7.1 \times 10^{-3}$ |
| 50 | 0.14 | $4.4 \times 10^{-2}$ |
| 100 | 0.35 | 0.11 |
| 200 | 0.64 | 0.21 |
| 500 | 0.99 | 0.37 |
| 550 | 1.00 | 0.39 |

Figure 1A Tunneling probability for pp-chain reactions as function of the energy: experimental vs simulated data (Rolfs and Rodney 1988)

## A.2 Example of the Energy dependence cross-section and $S(0)$ factor for $^3$He($\alpha$, $\gamma$)$^7$Be

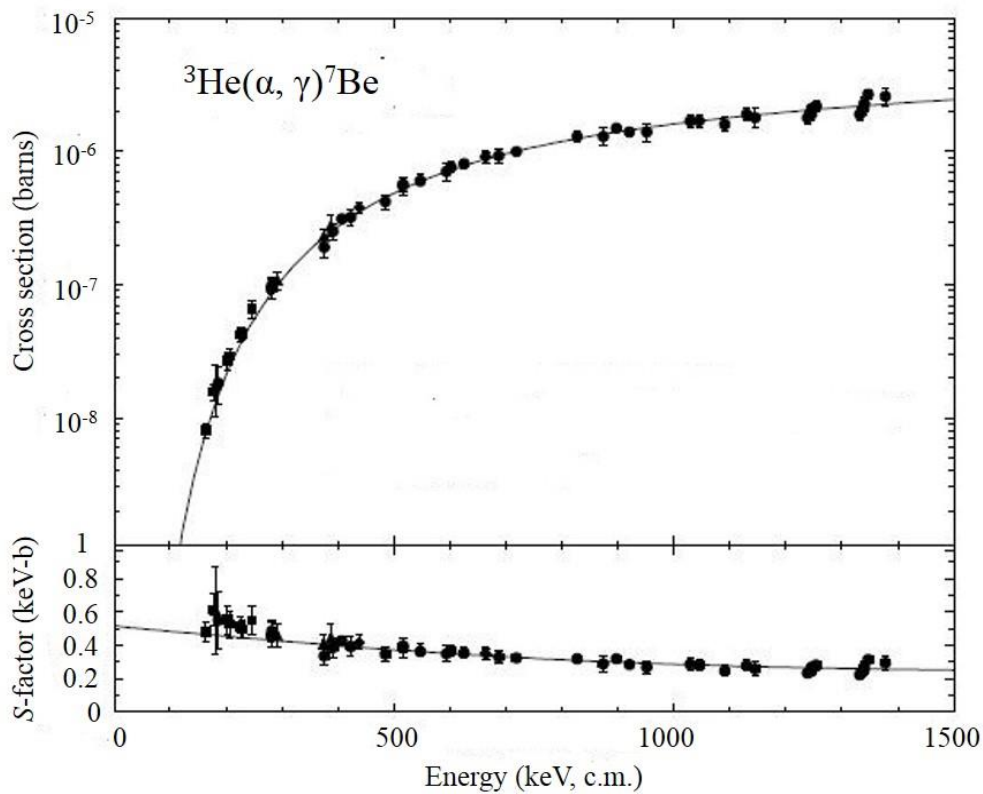

Figure 2A Astrophysical factor $S(E)$ and cross-section measurement of the $^3$H($\alpha$, $\gamma$)$^7$Be with error bars (Rolfs and Rodney 1988)

**A.3 Maxwell-Boltzmann velocity distribution**

At thermodynamic equilibrium the velocity of the nuclei related to the stellar gas can be described by the Maxwell Boltzmann Distribution:

$$\phi(v) = 4\pi v^2 \left(\frac{m}{2\pi kT}\right)^{3/2} exp\left[-\frac{mv^2}{2kT}\right] \tag{A1}$$

where the exponential term is related to the kinetic energy. The function $\phi(v)$ can be re-written:

$$\phi(v) \propto E^{1/2} exp[-E/kT] \tag{A2}$$

At low energy E<<kT the term $\phi(v)$ is linear dependant with the energy and has maximum vaue at E = kT. For nuclear reactions in stellar media the velocities for nuclei X and Y are decribed by the Maxwell Boltzmann Distribution velocity as:

$$\phi(v_{x,y}) = 4\pi v_{x,y}^2 \left(\frac{m_{x,y}}{2\pi kT}\right)^{3/2} exp\left[-\frac{m_{x,y}v_{x,y}^2}{2kT}\right] \tag{A3}$$

Hence the reaction rate per particle pair $\langle \sigma v \rangle$ can be found by a double integral over both velocity distributions with $v$ representing the relative velocity between the inracting nuclei:

$$\langle \sigma v \rangle = \iint_0^\infty \phi(v_x)\phi(v_y)\sigma(v)\,v\,dv_x dv_y \tag{A4}$$

In this equation vx and vy are respectively variables related to $v$ and $V$. Applying the reduced mass, the rate $\langle \sigma v \rangle$ can be described in terms of the variables $v$ and $V$ (center of mass velocity):

$$\langle \sigma v \rangle = \iint_0^\infty \phi(V)\phi(v)\sigma(v)\,v\,dV dv \tag{A5}$$

where the velocity related to $\phi(V)$ and $\phi(v)$ can be expressed as:

$$\phi(V) = 4\pi V^2 \left(\frac{M}{2\pi kT}\right)^{3/2} exp\left[-\frac{MV^2}{2kT}\right] \tag{A6}$$

$$\phi(v) = 4\pi v^2 \left(\frac{\mu}{2\pi kT}\right)^{3/2} exp\left[-\frac{\mu v^2}{2kT}\right] \tag{A7}$$

The nuclear cross-section $\sigma v$ is a function of the relative velocity between the interacting nuclei

$$\langle \sigma v \rangle = \int_0^\infty \phi(v)\sigma(v)v\,dv \tag{A8}$$

By inverting the equation above is possible to write:

$$\phi(v) = 4\pi v^2 \left(\frac{\mu}{2\pi kT}\right)^{3/2} \int_0^\infty v^3 \sigma(v) \, exp\left[-\frac{\mu v^2}{2kT}\right] dv \qquad \text{(A9)}$$

Inserting the centre-of-mass energy is possible to re-write the equation in the final form:

$$\langle \sigma v \rangle = \left(\frac{8}{\pi\mu}\right)^{1/2} \frac{1}{(kT)^{3/2}} \int_0^\infty \sigma(E) \, E \, exp\left(-\frac{E}{kT}\right) dE \qquad \text{(A10)}$$

# Appendix B

## B.1 Nuclear Reactions diagram

Representation of a nuclear reaction with entrance channel $0 + 1$ that proceed by an excited state in the compound nucleus and decay into particles $2 + 3$ of exit channel (Figure 1B):

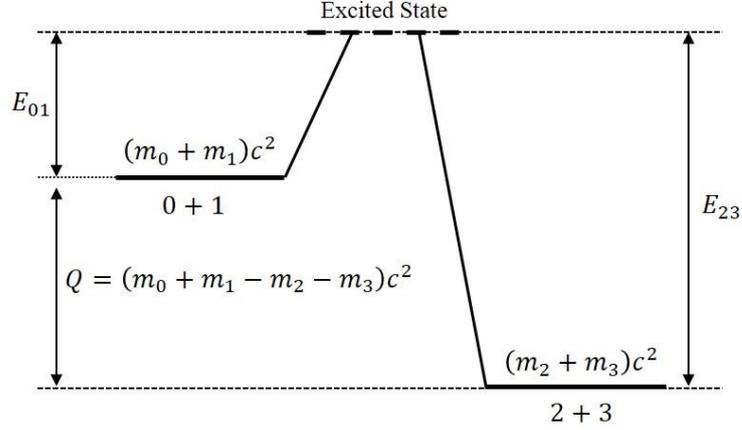

Figure 1B Schermatic representation of nuclear reaction with entrance, excited state and exit channel

In this nuclear reaction $0 + 1 \rightarrow 2 + 3$ with an excited state (black dashed-line) the cross-section can be expressed as:

$$\sigma_{01} = \pi \lambda_{01}^2 \frac{(2J + 1)}{(2J_0 + 1)(2J_1 + 1)} \qquad (B1)$$

where the indices $01$ describe the particles involved in entrance channel; the term $\pi \lambda_{01}^2$ is related to the quantum mechanic character of the cross section and $((2J + 1))/(2J_0 + 1)(2J_1 + 1)$ represents the quantised statistical factor $J$ related to the excited state for the compound nucleus while $J_0$ and $J_1$ is the spins of the nuclei 0 and 1 related to the entrance channel. From the quantum mechanical point of view the angular momentum $J$ has $2J+1$ magnetic substates. In some cases the reaction can proceed in opposite direction inverse reaction $2 + 3 \rightarrow 0 + 1$ and can be written as:

$$\sigma_{23} = \pi \lambda_{23}^2 \frac{(2J + 1)}{(2J_2 + 1)(2J_3 + 1)} \qquad (B2)$$

For these processes is possible to write an expression as the ratio of the two cross sections:

$$\frac{\sigma_{01}}{\sigma_{23}} = \frac{m_2 m_3 E_{23} (2J_2 + 1)(2J_3 + 1)(1 + \sigma_{01})}{m_0 m_1 E_{01} (2J_0 + 1)(2J_1 + 1)(1 + \sigma_{23})} \qquad (B3)$$

where $m_i$ and $E_i$ corresponds respectively to the mass and energy of the particles 0,1,2,3 for the entrance and exit channel, while $J_i$ are the quantum spin states of each particle.

# Appendix C

## C.1 Non Resonant reactions

In this process the projectile x on the plane wave is absorbed directly by the orbital shell of the final nucleus B. During that process a photon is emitted (Figure 1C):

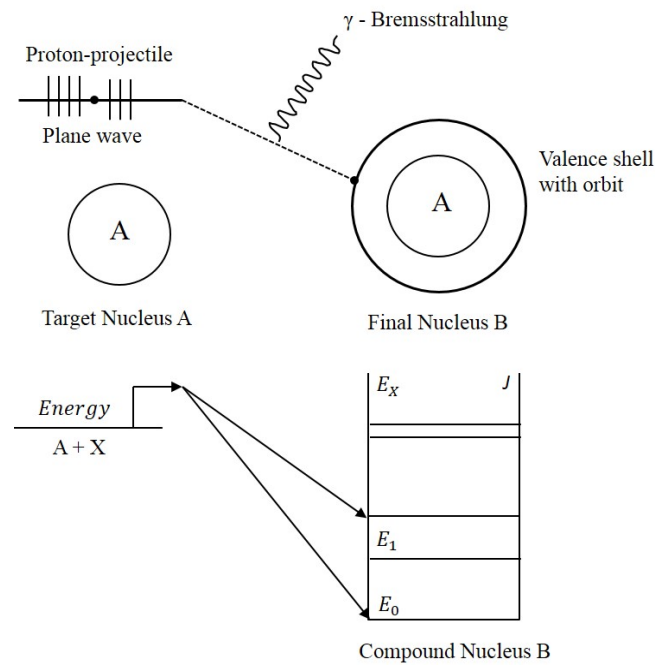

Figure 1C Direct capture non-resonant reaction process A(x, y)B with radiation emission

Since this process involves a single step is also called *direct capture* because there is a direct transition from the initial state to the final state. The direct capture includes an electromagnetic emission process called *Bremsstrahlung emission*. "Bremsstrahlung" is a german term indicating breaking radiation involving electrons that change velocity passing close to charged nuclei. This emission is not quantised, can generate continuous distribution in frequency and are also refered as free-free transitions because the upper and low level states are not bound.

This also a non-resonant process since it can involve all projectile quantised energy with cross-section and gradual variation of energy.

## C.2 Resonant Reactions

Another type of reaction can occu in the entrance channel when the nuclei $A + X$ react together to give an excited state $E_r$ of the compound nucleus that successively decays to a lower-lying state wih the emission of $\gamma$ radiation. This process occur only if the energy state of the entrance channel $Q + E_R$ matches the energy $E_r$ of the excited state in the compound nucleus as seen in Figure 2C (black bold line):

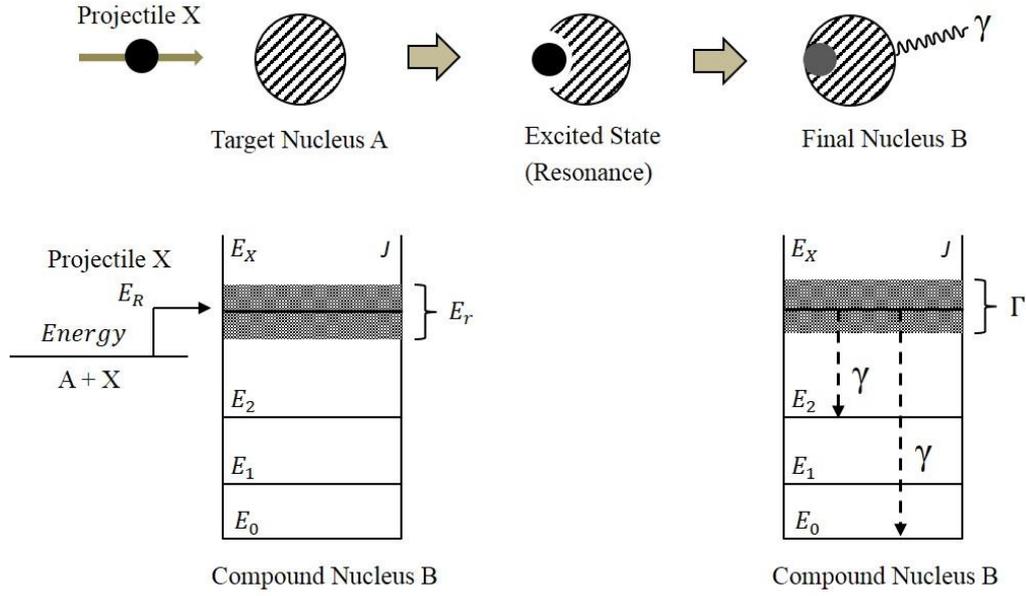

Figure 2C Capture Resonant reaction process A(x, y)B with decay radiation emission

This process is called a resonant reaction because it occurs only at fixed values of energy ($E_R$) of the projectile. If the excited state $E_r$ decays with emission of γ-rays to a lower energy state the cross-section can be decribed by a mathematical expression containing a term related to the formation of the compound and a term that referes to the γ decay of state Γ. Since the process involves a two step reaction, including an excited state is also called resonant reaction (i.e. triple α-process Chapter 3).

**C.3 Cross Section Equations for Resonant Reactions**

For a resonant reaction the cross section is given by Breit-Wigner expression indicated as:

$$\sigma(E) = \pi \lambda^2 \frac{(2J + 1)}{(2J_1 + 1)(2J_2 + 1)} \frac{\Gamma_1 \cdot \Gamma_2}{(E - E_R)^2 + (\Gamma/2)^2} \qquad (C1)$$

In this equation $\pi \lambda^2$ is the geometrical factor proportional to 1/E; $((2J + 1))/(2J_1 + 1)(2J_2 + 1)$ the spin factor ($J$ = spin state of the nuclei, $J_1$ = spin of the projectile; $J_2$ = spin of the target) and $\Gamma_1 \cdot \Gamma_2/(E - E_R)^2 + (\Gamma/2)^2$ is a term strongly energy dependent ($\Gamma_1$ = probability of compound formation via entrance channel; $\Gamma_2$ = probability of compound decay via exit channel; $\Gamma$ = total width of compound's excited state; $E_R$ = resonance energy). The Breit-Wigner equation is used for: fitting data to deduce resonance properties; in narrow resonance thermonuclear reaction rates; extrapolation of cross-sections when no measurement esist; experimental yields when resonace cannot be resolved.

## C.4 Narrow Resonance Case

Narrow resonances results from states that can only decay emitting a proton having an energy value much less than the Coulomb Barrier. As a consequence narrow resonance give an indication about of the internal nuclear structure and the lifetime involved. Conditions for narrow resonance are: it must be near relevant range $E_0$ to contribute to stellar rate; Maxwell Boltzmann distribution assumed constant over resonance region; partial widths also constant. Since the terms $E_R exp(-E/kT)$ have no changes $E = E_R$ they can be assumed constant and taken outside the integral:

$$\langle \sigma v \rangle = \left(\frac{8}{\pi\mu}\right)^{1/2} \frac{1}{(kT)^{3/2}} E_R \, exp\left(-\frac{E}{kT}\right) \int_0^\infty \sigma(E) \, dE \qquad \text{(C2)}$$

A schematic representation of the Maxwell-Boltzmann distribution and cross section for the narrow resonance is shown in Figure 3C:

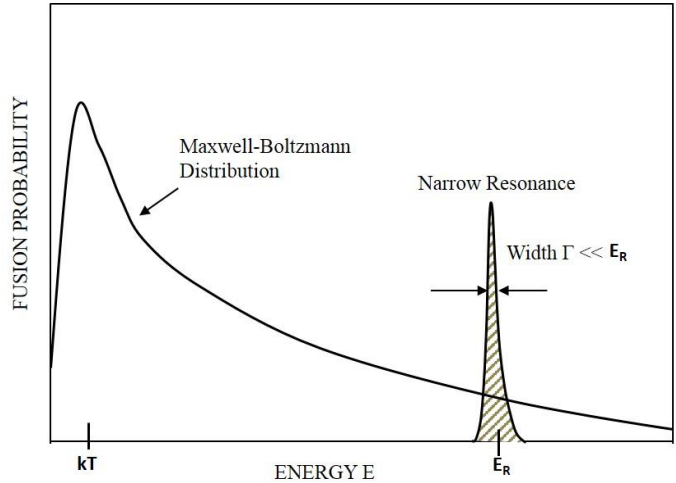

Figure 3C Maxwell-Boltzmann and cross-section schematic representation for a narrow resonance

Considering the Breit-Wigner cross section ($\sigma = \sigma_{BW}$) and $\lambda^2 \cong \lambda_R^2$ is possible to write:

$$\int_0^\infty \sigma_{BW}(E) \, dE = \pi \lambda_R^2 \omega \Gamma_\alpha \Gamma_b \int_0^\infty \frac{1}{(E - E_R)^2 + (\Gamma/2)^2} \, dE \qquad \text{(C3)}$$

The integral on the left-side with negligible energy dependence gives the following:

$$\int_0^\infty \sigma_{BW}(E) \, dE = 2\pi^2 \lambda_R^2 \omega \frac{\Gamma_a \Gamma_b}{\Gamma} \qquad \text{(C4)}$$

Where the $\omega$ corresponds to statistical factor indicated as:

$$\omega = \frac{(2J+1)}{(2J_1+1)(2J_2+1)} \tag{C5}$$

Combining the equations (C4) and (C5) and reminding that $\omega\gamma = \omega\Gamma_a\Gamma_b/\Gamma$ the reaction rate for a single narrow resonance cab be written as:

$$\langle\sigma v\rangle_{12} = \left(\frac{2\pi}{\mu kT}\right)^{3/2} \hbar^2 (\omega\gamma) \, exp\left(-\frac{E'_R}{kT}\right) \tag{C6}$$

where $\omega\gamma$ is also called the resonance strength:

$$\omega\gamma = \frac{(2J+1)}{(2J_1+1)(2J_2+1)} \frac{\Gamma_1 \cdot \Gamma_2}{\Gamma} \tag{C7}$$

In most cases:

$$\Gamma = \Gamma_1 + \Gamma_2 \tag{C8}$$

$$\Gamma_1 \ll \Gamma_2 \rightarrow \Gamma \approx \Gamma_2 \rightarrow \frac{\Gamma_1\Gamma_2}{\Gamma} \approx \Gamma_1 \tag{C9}$$

$$\Gamma_2 \ll \Gamma_1 \rightarrow \Gamma \approx \Gamma_1 \rightarrow \frac{\Gamma_1\Gamma_2}{\Gamma} \approx \Gamma_2 \tag{C10}$$

Therefore, the reaction rate is determined by the smaller width $\Gamma$.

# Appendix D LUNA Experiments

The proton-proton chain reaction studied at LUNA are represented in orange colour:

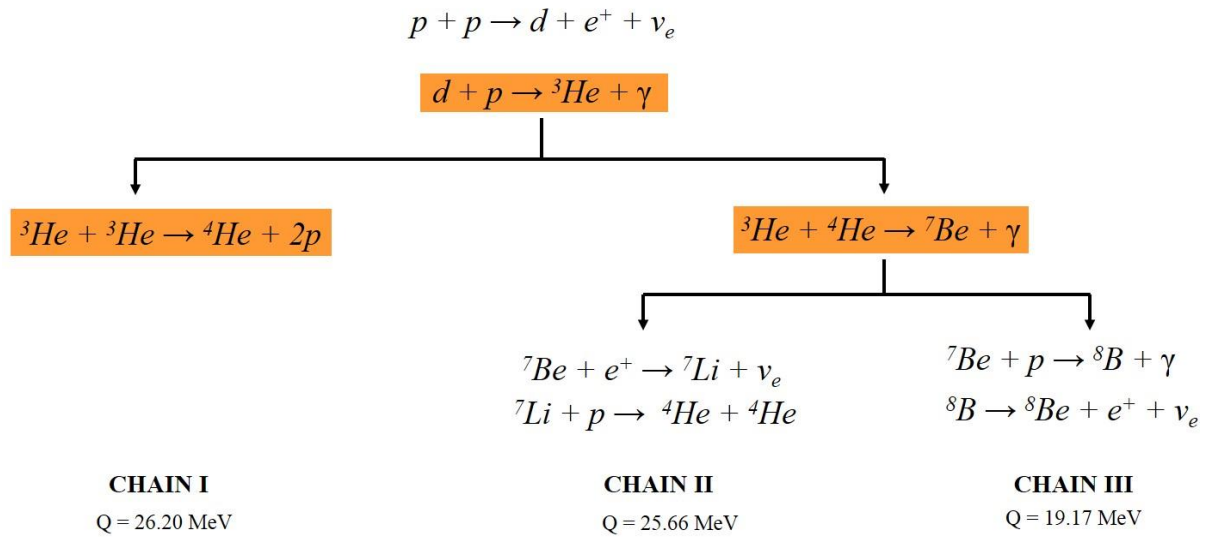

Figure 1D pp-Chain scheme (reactions studied at LUNA in orange colour)

In the sun the abundance of deuterium equilibrium is controlled by $^2$H(p, $\gamma$)$^3$He. This reaction play a key role in the life of protostar before enter in the main sequence stage (Figure 2D):

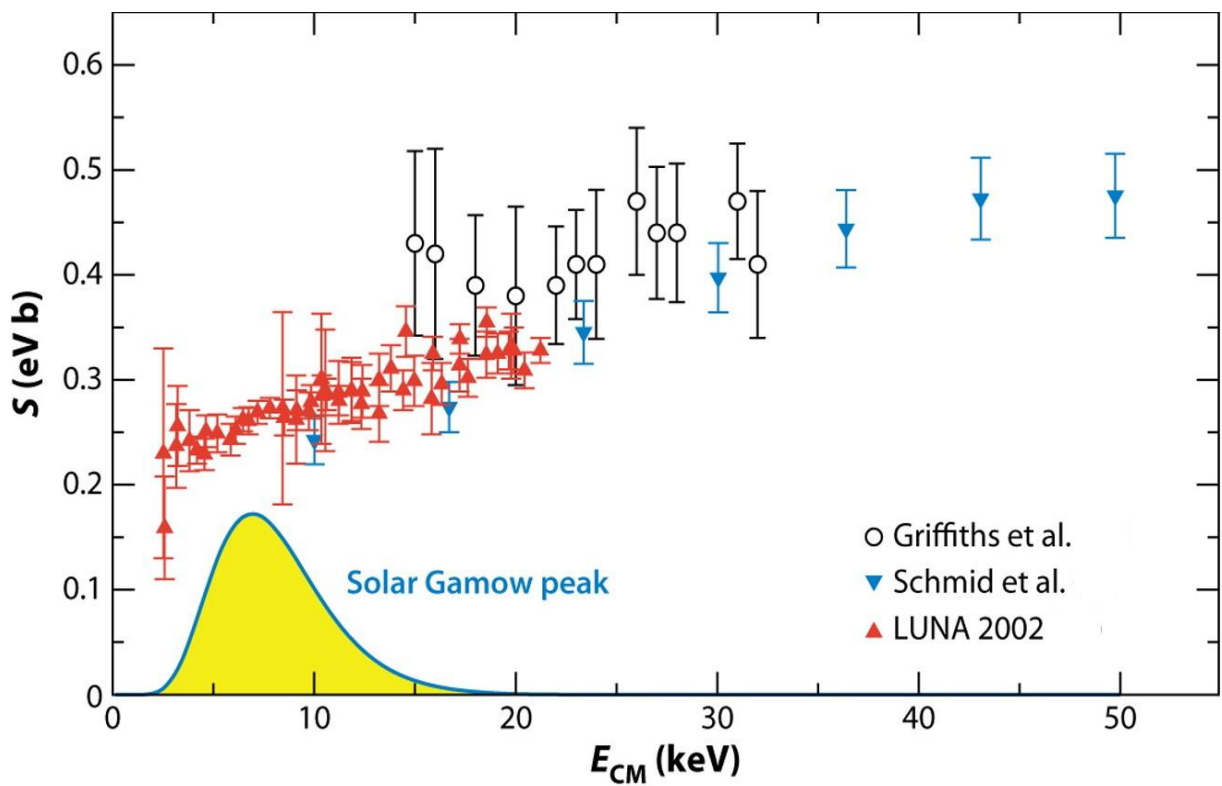

Figure 2D Astrophysical factor S($E$) measurement of the $^2$H(p, $\gamma$)$^3$He with error bars (Broggini et al. 2010)

At LUNA facilities the cross-section measurements have been obtained using 50-kV accelerator and the results in agreement with previous studies (Griffiths et al.; Smith et al.).

The cross-section measurement for $^3$He($^3$He, 2p)$^4$He with Gamow peak between 15 and 27 keV has been investigated at LUNA and is illustrated in Figure 3D:

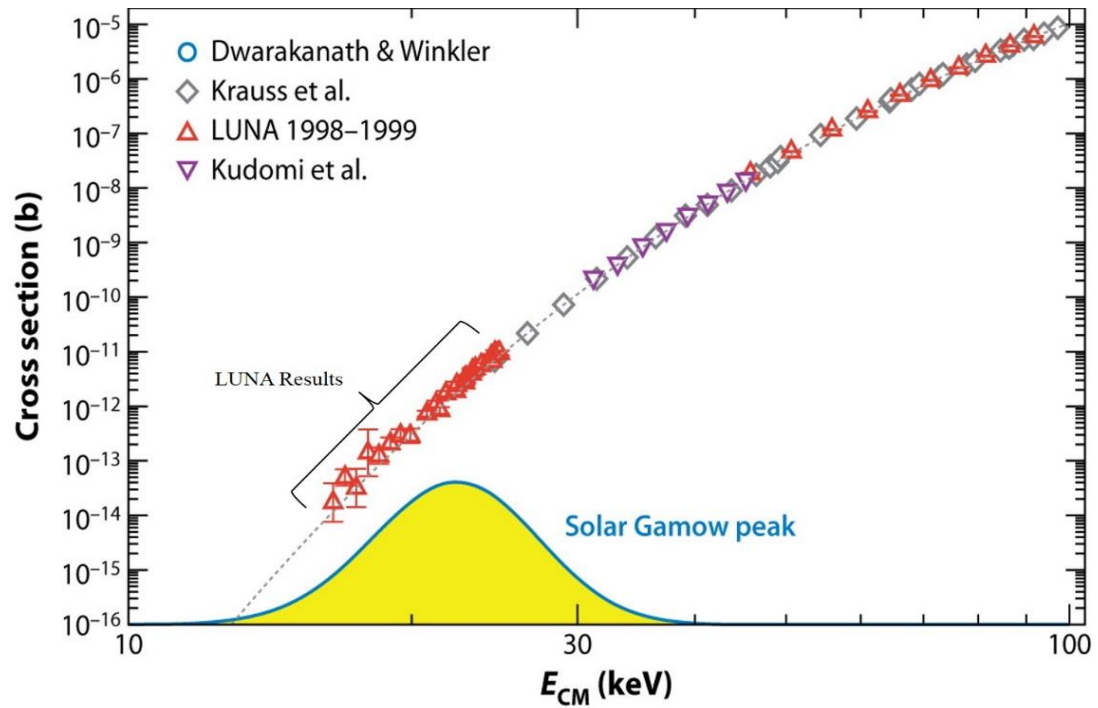

Figure 3D Cross section for $^3$He($^3$He, 2p)$^4$He reaction with dashes line related to the $S$(E) factor (Broggini et al. 2010 modified)

This study is extremely important because LUNA measurement (red triangles) have shown a cross section value of two order of magnitude smaller compared to previous studies (Krauss et al. 1987, grey romboids; Dwarakanath et al. 1971, blue circles; Kudomi et al. 2004, violet triangles). The spectrum for $^3$He($^4$He, γ)$^4$He reaction perfomed at energy beam of 250 keV is shown in Figure 4D:

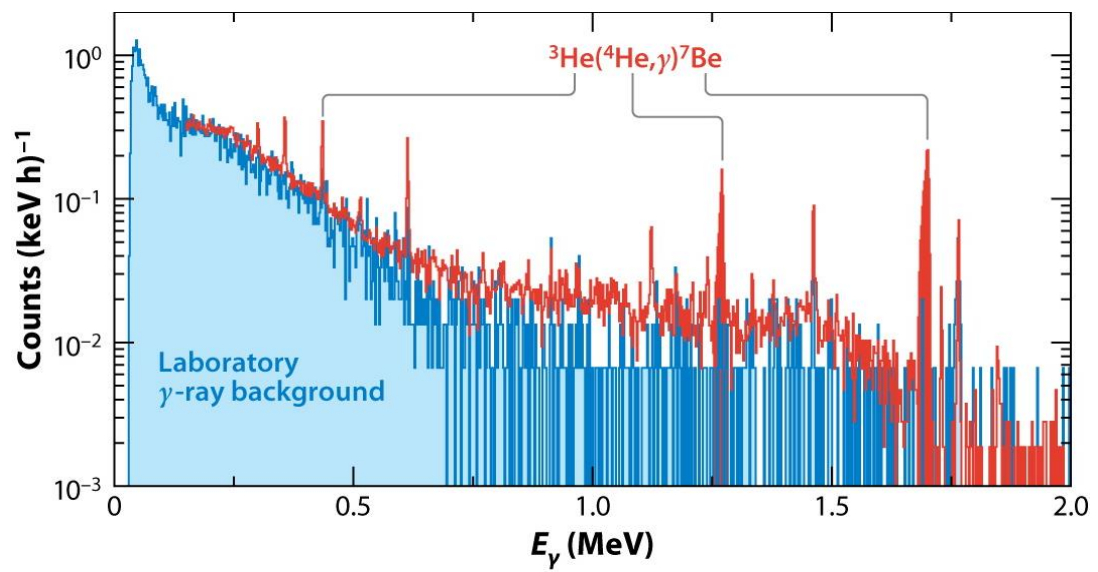

Figure 4D Spectrum for $^3$He($^4$He, γ)$^4$He reaction investigated at energy beam of 250 keV (Broggini et al. 2010)

Important measurements have been carried out at LUNA facilities (Figure 5D) using different detectors (red dashed-line; red triangles and red empty romboids) for the $^{14}N(p, \gamma)^{15}O$ reaction showing a twofold lower values than previous extrapolation results (Nacre, grey dashed-line; Lamb & Hester, blue quares and Schroder, brown triangles):

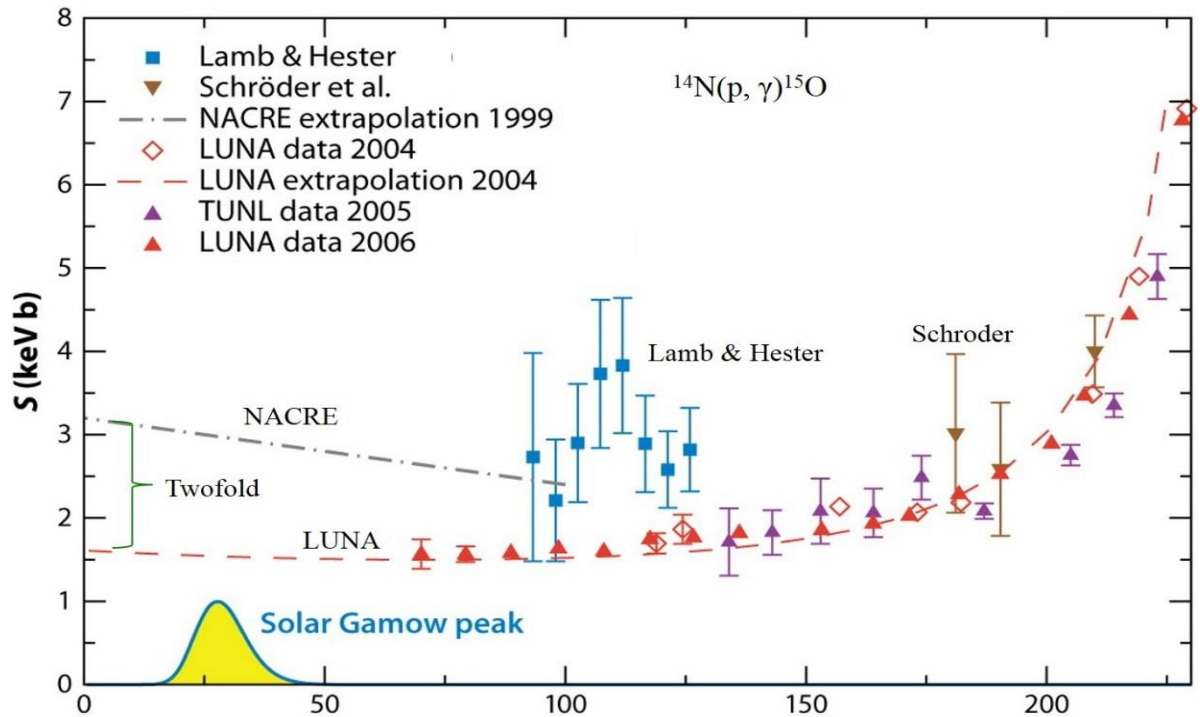

Figure 5D Astrophysical $S$(E) factor for $^{14}N(p, \gamma)^{15}O$ reaction with error bars (Broggini et al. 2010 modified)

As consequence of this result the CNO neutrino contribution in the Sun is lower and the globular cluster age increased by 1 to 14 billion years. The spectrum at 100 keV with cross section measurement is shown in Figure 6D:

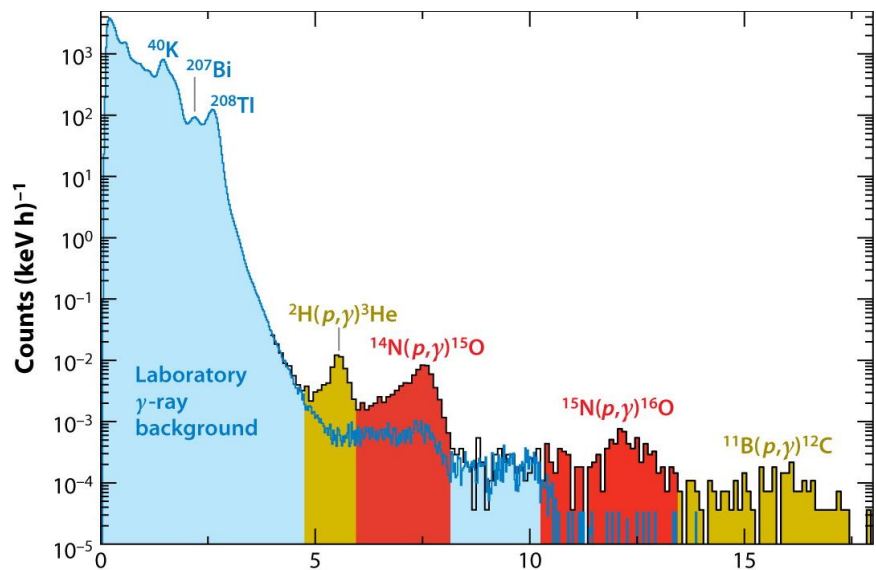

Figure 6D Spectrum using BGO detector: red colour  from $^{14}N(p, \gamma)^{15}O$ to $^{14}N(p, \gamma)^{16}O$; yellow and blue background (Broggini et al. 2010)